# Combining Transformer based Deep Reinforcement Learning with Black-Litterman Model for Portfolio Optimization

Ruoyu Sun[a], Angelos Stefanidis[b], Zhengyong Jiang[c,*], Jionglong Su[d,*]

[a] Xi'an Jiaotong-Liverpool University, School of Mathematics and Physics, Department of Financial and Actuarial Mathematics, Suzhou, 215123, China, Ruoyu.Sun19@student.xjtlu.edu.cn,
https://orcid.org/0009-0002-6052-0051

[b] Xi'an Jiaotong-Liverpool University Entrepreneur College (Taicang), School of AI and Advanced Computing, Suzhou, 215412, China, Angelos.Stefanidis@xjtlu.edu.cn,
https://orcid.org/0000-0002-4703-8765

[c] Xi'an Jiaotong-Liverpool University Entrepreneur College (Taicang), School of AI and Advanced Computing, Suzhou, 215412, China, Zhengyong.Jiang@xjtlu.edu.cn, +86 512 89167618,
https://orcid.org/0000-0001-8873-4073

[d] Xi'an Jiaotong-Liverpool University Entrepreneur College (Taicang), School of AI and Advanced Computing, Suzhou, 215412, China, Jionglong.Su@xjtlu.edu.cn, +86 512 88970778,
https://orcid.org/0000-0001-5360-6493

* Corresponding author.

**Abstract**

As a model-free algorithm, deep reinforcement learning (DRL) agent learns and makes decisions by interacting with the environment in an unsupervised way. In recent years, DRL algorithms have been widely applied by scholars for portfolio optimization in consecutive trading periods, since the DRL agent can dynamically adapt to market changes and does not rely on the specification of the joint dynamics across the assets. However, typical DRL agents for portfolio optimization cannot learn a policy that is aware of the dynamic correlation between portfolio asset returns. Since the dynamic correlations among portfolio assets are crucial in optimizing the portfolio, the lack of such knowledge makes it difficult for the DRL agent to maximize the return per unit of risk, especially when the target market permits short selling (i.e., the US stock market). In this research, we propose a hybrid portfolio optimization model combining the DRL agent and the Black-Litterman (BL) model to enable the DRL agent to learn the dynamic correlation between the portfolio asset returns and implement an efficacious long/short strategy based on the correlation. Essentially, the DRL agent is trained to learn the policy to apply the BL model to determine the target portfolio weights. In this model, we formulate a specific objective function based on the environment's reward function, which considers the return, risk, and transaction scale of the portfolio. Our DRL agent is trained by propagating the objective function's gradient to the policy function of our DRL agent. To test our DRL agent, we construct the portfolio based on all the Dow Jones Industrial Average constitute stocks. Empirical results of the experiments conducted on real-world United States stock market data demonstrate that our DRL agent significantly outperforms various comparison portfolio choice strategies and alternative DRL frameworks by at least 42% in terms of accumulated return. In terms of the return per unit of risk, our DRL agent significantly outperforms various comparative portfolio choice strategies and alternative strategies based on other machine learning frameworks.

**Statements and Declarations**

The authors did not receive support from any organization for the submitted work. The authors have no relevant financial or non-financial interests to disclose.

# 1. Introduction

Portfolio optimization refers to the process of constructing an optimal portfolio based on an investor's objectives in terms of return and risk. In portfolio optimization, we seek to allocate the funds into a given set of assets to maximize the return and control the risk during continuous trading periods. The first portfolio optimization model, i.e., the mean-variance model, was proposed by Markowitz [1]. This model provided a rigorous operational theory for the portfolio optimization problem [2]. Contemporary academic research on portfolio optimization is influenced mainly by Markowitz's fundamental theory [3], which advocates that investors should rigorously exploit the dynamic correlation between the returns of the portfolio asset and implement the long/short strategy to maximize a specified objective function. Typically, the objective function is defined as the expected return penalized by the risk of the portfolio [4].

Advancement in artificial intelligence (AI) brings new insights into financial technology. In recent years, various AI techniques have been utilized by investors to assist in optimizing their investment portfolios. These techniques include neurodynamic optimization [5-8], market sentiment analysis based on natural language processing techniques [9], deep reinforcement learning algorithms [10], and stock price forecasting [11-13]. In these techniques, deep reinforcement learning (DRL) algorithms have shown promising results in portfolio optimization in consecutive trading periods [14]. DRL agent learns and makes decisions by interacting with the environment in an unsupervised way. As a model-free algorithm, it does not rely on the specification of the joint dynamics across the portfolio assets. It guarantees that the DRL agent can dynamically adapt to market changes [15]. However, in the common practice of applying the DRL algorithm for portfolio optimization, a DRL agent is trained to apply a policy neural network to determine the target portfolio weights based on the current environmental state. The policy networks of these agents are trained based on the environment's reward function, which is constructed based on the return and the corresponding risk of the portfolio. In this framework, it is difficult for these DRL agents to learn a policy that can take into account of the dynamic correlation between the portfolio asset returns. Fully exploiting the correlation of the portfolio asset returns is crucial for the long/short strategy. By utilizing the specific co-movement patterns, assets with negative correlation may offset each other's performance, with one asset's gains potentially counterbalancing the losses of another during market fluctuations. This diversification approach may stabilize the overall portfolio's returns by mitigating unsystematic risk attributable to individual stocks' volatility [16]. By learning the correlation between the portfolio asset returns, the investors can improve the return per unit of risk by implementing a long/short strategy based on the correlation. Since the DRL agent can not implement an efficacious long/short strategy based on the correlation between the portfolio asset returns, it is difficult for traditional DRL agents to maximize the return per unit of risk, especially if the market permits short selling.

To tackle this challenge, we first consider training the DRL agent to learn the policy based on Markowitz's mean-variance model when determining the target portfolio in the market permit short selling. In Markowitz's mean-variance model, the correlation among each portfolio asset can be taken into account when calculating the target portfolio weights. Unfortunately, Markowitz's mean-variance model may lead to the issue of error maximization [17]. As a result, the optimized portfolio based on Markowitz's mean-variance model overweights the assets with large estimation errors [18], which may lead to poor out-sample performance of the DRL agent's policy [110]. In response to the limitation of Markowitz's mean-variance model, the DRL agent is trained to apply the mathematical Black-Litterman (BL) model to make portfolio choices. BL model is a Bayesian model that can combine the DRL agent's subjective views of the expected return of the assets in the portfolio and the prior distribution of asset returns based on historical data. Lee [19] proved that the BL model can efficiently overcome the issue of error-maximization [17]. Specifically, the DRL agent is trained to learn a policy function to apply the mathematical BL model to determine the portfolio weights, which is termed as the Black-Litterman model based deep reinforcement learning agent (BDA) in our paper. In this way, the DRL agent can learn the dynamic correlation between the portfolio asset returns and implement a long/short strategy based on the correlation. Given that the time series data of financial asset prices are nonlinear, dynamic and chaotic [21] and the neural network is an effective tool to recognize nonlinear patterns, we train the DRL agent to apply the neural network to recognize the nonlinear patterns [22]. Specifically, the DRL agent applies a Transformer neural network [20] and a convolutional neural network to output the return expectation and risk aversion in the BL model. The prior distribution parameters in the BL model are calculated based on historical data. In the Transformer network, we remove the position encoding module to mitigate overfitting and to realize the generalization ability of the policy function. By removing the position encoding module, we let our BDA concentrate on learning the nonlinear correlation from multiple concurrent series of portfolio asset returns when determining the subjective view for the expected return.

As the DRL agent engages in portfolio optimization, its action space is continuous and high-dimensional. Since there is no restriction on the weights of investment allocations for each portfolio asset, the computation cost for the DRL agent to fully explore the action space is prohibitively large. In traditional actor-critic algorithms with deterministic policies designed for portfolio optimization (i.e., Deep Deterministic Policy Gradient (DDPG) [33] and Twin Delayed DDPG (TD3) [35]), agents face substantial challenges in exhaustively exploring the action space. Concurrently, the training of the critic network is confronted with the issue of the curse of dimensionality. To effectively train our agent, we adopt the deterministic policy gradient algorithm of Jiang et al. [23] to train our agent. In the training process, we directly formulate the objective function based on the environment's reward function and propagate the objective function's analytic gradients back into the policy function of BDA.

The motivation of our research is two-fold. First, although DRL agents have shown promising results in portfolio optimization, it remains a challenge for these DRL agents to learn a policy that is aware of the dynamic correlation between the portfolio asset returns. In scenarios where the market permits short selling, it is difficult for them to improve the return per unit of risk by utilizing the dynamic correlation between portfolio asset returns and implementing an efficacious long/short strategy. Hence, we plan to overcome this issue to improve the DRL agent's performance in return per unit of risk when optimizing the portfolio in consecutive trading periods. Second, Black and Litterman propose a mean-variance model, which can incorporate the investors' views regarding the return expectations of the portfolio assets with different confidence levels based on the prior distribution to derive a posterior distribution for portfolio optimization. Scholars [19] proved that the BL model can efficiently overcome the issue of error-maximization [17] in Markowitz's mean-variance model. As a Bayesian model, the BL model can combine the DRL agent's subjective views of the expected return of some of the assets in the portfolio and the prior distribution of asset returns based on historical data. Consequently, the DRL agent can make portfolio optimization decisions by utilizing the BL model based on its subjective views of return expectation and risk aversion. Since the BL model is a mean-variance model, our DRL agent can implement a long/short strategy based on the dynamic correlation between portfolio asset returns when determining the target portfolio weights.

The key contributions of our work are summarized below:
- To the best of our knowledge, we are the first to train the DRL agent to learn the policy in applying the BL model for portfolio optimization in the target market, allowing short-selling. Consequently, our DRL agent can learn the dynamic correlation between the portfolio asset returns and implement a long/short strategy according to the correlation to improve the return per unit of risk. When testing our BDA based on the data in the US stock market, the empirical results show that, within the US stock market, the long/short strategy of our DRL agent can significantly outperform other comparative strategies in both accumulative return and return per unit of risk. The empirical results of our experiments demonstrate the feasibility of learning the policy in applying the BL model to improve the return per unit of risk while maintaining outstanding performance in accumulated return.
- We find that the well-training of the critic network is particularly challenging due to the "curse of dimensionality". To overcome this, we formulate the training objective function based on the environment's reward function and directly propagate the objective function's analytic gradients back into the DRL agent's policy function. The empirical results in the ablation study prove that such a method can ensure that the DRL agent can efficiently maximize the accumulative reward within the training environment.
- To realize the generalization ability of the policy, our DRL agent employs a modified Transformer network stripped of the position encoding module to learn the policy for determining the subjective view of the expected return. This approach allows the DRL agent to concentrate on learning the nonlinear correlation between the portfolio asset returns when determining the subjective view for the expected return. Empirical results indicate that such a learning pattern can mitigate overfitting and realize the out-sample generalization ability of our DRL agent's policy.
- Our research suggest that, in the unconstrained portfolio optimization problem in consecutive trading periods, establishing training target value for the function derived from the environment's reward function instead of maximizing such function in the training process is a feasible way to effectively avoid overfitting and realize the generalization ability of the DRL agent's policy in the out-sample environments.

The rest of the paper is organized as follows. Section 2 presents the literature review. Section 3 gives an overview of the mathematical formulas in the periodical portfolio optimization problem and defines the objective of the DRL agent. Section 4 details the modules in the framework of the DRL method. Section 5 introduces the topology of the neural networks applied during the decision process of the DRL agent. Section 6 gives the detailed experimental design and the empirical results of each experiment. Finally, Section 7 gives the conclusions and future work.

## 2. Literature review

In recent years, researchers have applied different DRL algorithms [24] to train the agent for portfolio optimization in consecutive trading periods. In these attempts, the applied DRL algorithms can be divided into policy-based algorithms and value-based algorithms according to their training method. In the application of the policy-based algorithms for portfolio optimization, the DRL agents are all trained to learn a policy to output the action for portfolio optimization. In the policies of these DRL agents, there are attempts at multiple decision logic. One of the most dominant decision logic is training the DRL agent to score the portfolio assets according to the specific reward function.

Subsequently, the portfolio weights are transformed from the asset scores by the softmax function. Several research studies have been conducted to explore the application of the policy network with such decision logic. Jiang et al. [23] propose the Ensemble of Identical Independent Evaluators (EIIE) policy topology based on such decision policy and apply the Deterministic Policy Gradient (DPG) algorithm to train the DRL agent's policy. Based on the EIIE framework of Jiang et al. [24][23], Sun et al. [25] propose to apply the deep residual shrinkage neural network to function as the identical independent evaluator in Jiang et al.'s [23] EIIE framework to optimize the policy function of the DRL agent. Shi et al. [26] proposed a novel neural network topology named Ensemble of Identical Independent Inception (EI$^3$) to enable the DRL agent to analyze the multi-sale price movement information. Song et al. [27] propose the Stochastic Policy with Distributional Q-network (SPDQ) by integrating the Soft actor-critic (SAC) [28] algorithm with Quantile-Regression DQN (QR-DQN) [29]. SARL [30] is a state-augmented DRL framework for portfolio optimization. In its framework, diverse information is leveraged to make asset price predictions to augment the state of the DRL agent. GPM [15] is a multi-scale graph convolutional network based DRL framework for portfolio optimization. It takes full account of the temporal and relational features of the portfolio by utilizing relational graph convolutions and multi-scale convolutions. DeepTrader, proposed by Wang et al. [32], is a DRL framework consisting of an asset scoring unit and a market scoring unit. In their framework, an asset scoring unit is applied to score the assets by analyzing each portfolio asset's rise in the future and the interrelationship among all assets. A market scoring unit is applied to leverage the market sentiment indicators to analyze the financial situation.

Some researchers propose to train the DRL agent to output the target amount of buying or selling shares on each asset in the portfolio by applying the policy network. As for the research studies in exploring the application of the policy network with such decision logic, FinRL, proposed by Liu et al. [32], is a framework in which several off-the-shelf DRL algorithms are applied to train the DRL agent for portfolio optimization. The DRL algorithms in the FinRL framework include deep deterministic policy gradient (DDPG) [33], twin delayed deep deterministic policy gradient (TD3) [35], proximal policy optimization (PPO) [34], advantage actor critic (A2C) [62], and soft actor-critic (SAC) [28]. Based on the SAC implementation in the framework of FinRL [32], Gao et al. propose StockFormer [84], a hybrid model that can combine the benefits of DRL agents in terms of policy flexibility with the forward modelling powers of predictive coding.

Among the portfolio optimization research on the value-based algorithm, the DRL agents are all trained to directly score the specific portfolio management actions in the action space. In the decision logic of these DRL agents, the agents are trained to learn a value function to score each portfolio optimization action in the action space and choose the action with the highest score. As for the research studies in exploring the application of such value function, Gao et al. [36] construct a hierarchical deep Q-learning framework for portfolio optimization. In their framework, portfolio assets are assigned to different Deep Q-Network (DQN) to reduce the action number and improve the functionality of the algorithm. Shavandi and Khedmati [37] design a multi-agent deep Q-learning framework where DRL agents are trained to learn trading in different timeframes. Lucarelli and Borrotti [38] propose two deep Q-learning models based on Dueling Double Deep Q-network (DD-DQN) [39] and Double Deep Q-Network (D-DQN) [40] for portfolio optimization in the Cryptocurrency market. Based on the deep Q-learning model in [41], Lucarelli and Borrotti further extend the model by combining the deep Q-learning technique with the multi-agent framework for portfolio optimization

## 3. Preliminaries

This section defines the basic financial concepts used throughout the paper. Subsequently, it also describes the trading process during each trading period and defines the periodic portfolio management problem we set out to address.

### 3.1 Basic Financial Concepts

**Definition 3.1.1 (Trading period and Observation period)** A trading period is the minimum time unit used to reallocate the fund. The timeline is divided into several trading periods with equal lengths. As given in Figure 1, the $t^{\text{th}}$ trading period is defined as the period interval $(t, t+1]$, $t = 0, 1, 2, \ldots, T_f - 1$, where $T_f$ is the total number of trading periods. The portfolio weights are determined at the end of each trading period. According to Li et al. [91], high-frequency stock trading data always exhibits a low signal-to-noise ratio. To prevent the DRL agent from learning too much noisy information from the environment, we extend the trading period to decrease the trading frequency. Here, the trading period is taken to be five trading days. The $k^{\text{th}}$ trading day within the $t^{\text{th}}$ trading period is denoted as $t_k$, $k = 1, 2, .., 5$.

The observation period is the minimum time to observe the price indices. To better describe the state of the environment, the frequency of collecting data is set to be higher than the trading frequency. We observe the price indices at the end of each trading day. Hence, the price indices will be observed five times in each trading period.

**Definition 3.1.2 (Long position and short position)** In portfolio optimization, taking a long position [42] refers to the purchase of the target stock with the expectation that the stock will rise in value. If the stock price increases in future, the investor can gain from any increase in the stock price.

In portfolio optimization, taking a short position is also known as short selling [43]. It involves borrowing shares of the target stock and immediately selling them in the open market with the intention of buying them back later at a lower price. If the stock price subsequently decreases, the investors can buy back the shares at a lower price and return the shares to the lender. The investor's profit is the difference in the price of the asset between the time of the initial sale and the repurchase.

**Definition 3.1.3 (Target Portfolio weights)** The portfolio weights vector $w_t$ represents the target portfolio weights at the beginning of the $t^{\text{th}}$ trading period, i.e.,

$$w_t = [w_{1,t}, w_{2,t}, \ldots, w_{n,t}]^T,$$

where the $i^{\text{th}}$ component $w_{i,t}$ represents the proportion of the total portfolio value (money) invested in asset $i$ at the beginning of the $t^{\text{th}}$ trading period, and $n$ is the number of assets in the portfolio. Furthermore, $w_{0,t} = 1 - \sum_{i=1}^{n} w_{i,t}$ is defined as the target weight of the risk-free asset (cash) at the end of the $t^{\text{th}}$ period. In this research, to achieve the target portfolio weights $w_t$, the asset reallocation is carried out at the end of each trading period. Hence, the target portfolio weights $w_t$ should be determined at the end of the $t - 1^{\text{th}}$ trading period.

**Definition 3.1.4 (Price vector)** The price vector $p_t$ is the adjusted closing price vector of the last trading day within the $t^{\text{th}}$ trading period, i.e.,

$$p_t = [p_{1,t}, p_{2,t}, \ldots, p_{n,t}]^T,$$

where $p_{i,t}$ is the adjusted closing price of asset $i$ at the last trading day within the $t^{\text{th}}$ trading period. The adjusted closing price refers to the finalized trading price of a stock that has undergone adjustments to account for various corporate actions such as dividends, stock splits, and other significant events. Hence, it is an essential index in financial analysis and investment decision-making. The transactions executed at the end of the $t^{\text{th}}$ trading period are based on the price vector $p_t$. As given in Figure 1, the adjusted closing price of asset $i$ at the $k^{\text{th}}$ trading day within the $t^{\text{th}}$ trading period is defined as $p_{i,t_k}$. Hence, the corresponding price vector of the portfolio is

$$p_{t_k} = [p_{1,t_k}, p_{2,t_k}, \ldots, p_{n,t_k}]^T.$$

Noticeably, in each trading period $t$, the adjusted closing price vector on the fifth trading day $p_{t_5}$ is also the price vector $p_t$ of this trading period.

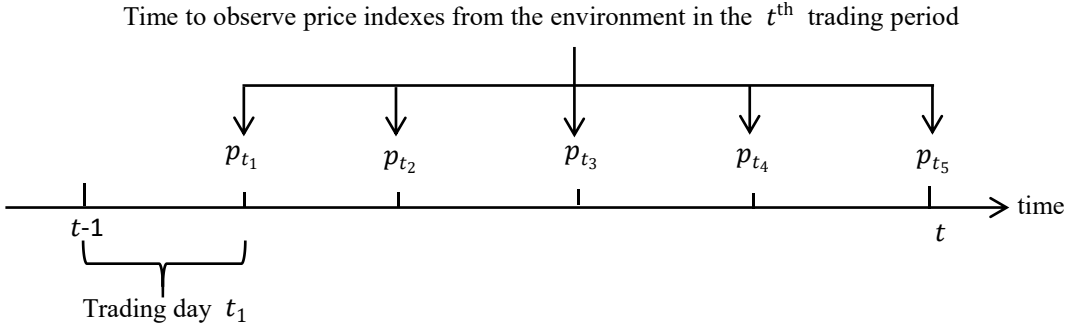

Fig 1. The observation times scheduled in the $t^{\text{th}}$ trading period. There are five trading days $t_k, k = 1, 2, \ldots, 5$ in the $t^{\text{th}}$ trading period. At the end of each trading day $t_k, k = 1, 2, \ldots, 5$, we collect the current price vector $p_{t_k}, k = 1, 2, \ldots, 5$.

**Definition 3.1.5 (Return vector).** Let $R_t$ denote the assets return vector in the $t^{\text{th}}$ trading period, which is formulated as:

$$R_t = \left[\log_2\left(\frac{p_{1,t}}{p_{1,t-1}}\right), \log_2\left(\frac{p_{2,t}}{p_{2,t-1}}\right), \ldots, \log_2\left(\frac{p_{n,t}}{p_{n,t-1}}\right)\right]^T$$
$$= [R_{1,t}, R_{2,t}, \ldots, R_{n,t}]^T, \tag{1}$$

In Equation (1), $R_{i,t}$ is the logarithmic return of asset $i$ at the $t^{\text{th}}$ trading period. To further describe the return distribution within the $t^{\text{th}}$ trading period, we define the logarithmic return of asset $i$ at the $k^{\text{th}}$ trading day within the $t^{\text{th}}$ trading period as $z_{i,t_k}$, i.e.,

$$z_{i,t_k} = \begin{cases} \log_2\left(\frac{p_{i,t_k}}{p_{i,t_{k-1}}}\right) & k = 2, 3, 4, 5 \\ \log_2\left(\frac{p_{i,t_k}}{p_{i,t-1}}\right) & k = 1 \end{cases}.$$

Hence, the corresponding return vector of the portfolio at the $k^{\text{th}}$ trading day within the $t^{\text{th}}$ trading period is defined as:

$$z_{t_k} = \begin{cases} \left[\log_2\left(\frac{p_{1,t_k}}{p_{1,t_{k-1}}}\right), \log_2\left(\frac{p_{2,t_k}}{p_{2,t_{k-1}}}\right), \ldots, \log_2\left(\frac{p_{n,t_k}}{p_{n,t_{k-1}}}\right)\right]^T & k = 2, 3, 4, 5 \\ \left[\log_2\left(\frac{p_{1,t_k}}{p_{1,t-1}}\right), \log_2\left(\frac{p_{2,t_k}}{p_{2,t-1}}\right), \ldots, \log_2\left(\frac{p_{n,t_k}}{p_{n,t-1}}\right)\right]^T & k = 1 \end{cases}$$

The expectation and the standard deviation of the return $z_{i,t_k}$ are denoted as $\mu_{i,t}$ and $\sigma_{i,t}$. For $i \neq j$, the correlation coefficient of $z_{i,t_k}$ and $z_{j,t_k}$ is represented by $\rho_{ij}$, and the expectation of $z_{t_k}$ can be denoted as

$$\mu_t = [\mu_{1,t}, \mu_{2,t}, \ldots, \mu_{n,t}]^T \tag{2}$$

and the symmetric covariance matrix of $z_{t_k}$ can be denoted as

$$\Sigma_t = \begin{bmatrix} \sigma_{11,t}^2 & \cdots & \sigma_{1n,t}^2 \\ \vdots & \ddots & \vdots \\ \sigma_{n1,t}^2 & \cdots & \sigma_{nn,t}^2 \end{bmatrix}, \tag{3}$$

where $\sigma_{ii,t}^2$ is the variance of asset $i$ at the $t^{th}$ trading period and $\sigma_{ij,t}^2$ is the covariance between asset $i$ and asset $j$ at the $t^{th}$ trading period, $i \neq j$.

**Definition 3.1.6 (Target quantity).** The target quantity $q_t = [q_{1,t}, \ldots, q_{n,t}]^T$ is the target position sizing in the different portfolio stocks at the $t^{th}$ trading period, which is calculated based on the target portfolio weights $w_t$ determined at the end of the $t-1^{th}$ trading period and the total investment amount $T_t$:

$$q_t = [q_{1,t}, q_{2,t}, \ldots, q_{n,t}]^T = \lfloor (T_t w_t) \oslash p_{t-1} \rfloor, \tag{4}$$

where $\oslash$ is the element-wise division. In Equation (4), we apply the floor function so that the target quantity $q_{i,t}$ contains integers since we assume that one share of stock cannot be divided. In the long/short strategy, the leverage transaction makes investors trade larger position sizing with the same principal sum. To reduce the transaction cost and lending cost caused by using leverage, we control the total size of the long and short positions in the portfolio by limiting the investment amount in each trading period. The total investment amount $T_t$ in each trading period is set as half of the initial amount $T_1$, i.e.,

$$T_t = 0.5 T_1, \text{ for } t = 1, \ldots, T.$$

This means that the value of the total assets is different from the value of the portfolio in each trading period.

The market order is the number of stock shares we buy or sell at the end of each trading period, which is the difference in target quantities between two adjacent trading periods. The market order vector $\Delta q_t$ at the $t^{th}$ trading period is defined as:

$$\Delta q_t = [\Delta q_{1,t}, \Delta q_{2,t}, \ldots, \Delta q_{n,t}]^T = q_t - q_{t-1}.$$

**Definition 3.1.7 (Risk-free asset value)** We define $c_t$ to be the value of the risk-free asset (cash) within the $t^{th}$ trading period. When taking into account of the transaction costs incurred by reallocating the assets, the value of cash $c_t$ can be depicted as:

$$c_t = \begin{cases} c_{t-1} - \Delta q_t^T p_{t-1} - r_s |\min(q_{t-1}, \mathbf{0})^T| p_{t-1} - \alpha |\Delta q_t|^T p_{t-1} & c_{t-1} \geq 0 \\ c_{t-1}(1 + r_l) - \Delta q_t^T p_{t-1} - r_s |\min(q_{t-1}, \mathbf{0})^T| p_{t-1} - \alpha |\Delta q_t|^T p_{t-1} & c_{t-1} < 0 \end{cases}, \tag{5}$$

where $\mathbf{0}$ is a $n \times 1$ zero vector. In Equation (5), $\alpha$, $r_l$ and $r_s$, respectively, represent the commission fee rate, lending rate of cash, and lending rate of stock. In Equation (5), the terms $\alpha |\Delta q_t| p_t$ and $r_s |Min(q_{t-1}, 0)^T| p_{t-1}$, respectively, denote the trading cost and the cost of the short position in this trading period. In recent years, the risk-free rate (represented by the U.S. Treasury rate) has exhibited instability. In some periods (i.e., 2020 and 2021, the risk-free rate in the American market is close to zero. Hence, we do not consider the risk-free benefit that the risk-free asset can obtain from the market. The risk-free rate is assumed to be zero in our research.

**Definition 3.1.8 (Total asset value).** Total asset value is the total value of all the assets. We define $v_t$ as the total asset value before the target portfolio weights $w_{t+1}$ is executed at the end of the $t^{th}$ trading period, i.e.,

$$v_t = c_t + q_t^T p_t,$$

where the term $q_t^T p_t$ represents the total value of the stocks held at the end of $t^{th}$ trading period. To further describe the changes in the value of total assets, we define the value of total assets at the end of the $k^{th}$ trading day within the $t^{th}$ trading period as $v_{t_k}$. Hence, the value of total assets at the end of the fifth trading period $v_{t_5}$ is the total asset value $v_t$ of this trading period.

**Definition 3.1.9 (Logarithmic return)** The logarithmic return of the portfolio in the $t^{th}$ trading period is defined as $\xi_t$, and it can be calculated as:

$$\xi_t = \log_2 \left( \frac{v_t - v_{t-1}}{T_t} + 1 \right), \tag{6}$$

and the daily logarithmic return of the portfolio at the $k^{th}$ trading day within the $t^{th}$ period is

$$\vartheta_{t_k} = \begin{cases} \log_2 \left( \dfrac{v_{t_k} - v_{t-1} + T_t}{v_{t_{k-1}} - v_{t-1} + T_t} \right) & k = 2,3,4,5 \\ \log_2 \left( \dfrac{v_{t_k} - v_{t-1}}{T_t} + 1 \right) & k = 1 \end{cases}.$$

**Definition 3.1.10 (Variance)** The variance of the portfolio $V_t$ measures the risk of the portfolio at time $t$. The variance of the portfolio is calculated as:

$$V_t = w_t^T \Sigma_t w_t. \tag{7}$$

where $\Sigma_t$ is the covariance matrix of the portfolio return vector defined in Equation (3). In our model, the covariance matrix $\Sigma_t$ is calculated based on the return distribution from the current trading period and the previous $m-1$ trading periods:

$$\Sigma_t = \frac{\left[ X_t - \frac{1}{5m} \sum_{i=0}^{m-1} \sum_{k=1}^{5} z_{t-i_k} \right] \left[ X_t - \frac{1}{5m} \sum_{i=0}^{m-1} \sum_{k=1}^{5} z_{t-i_k} \right]^T}{5m - n - 1},$$

where $z_{t-i_k}$ is the return vector of the portfolio assets at the $k^{th}$ trading day within the $t-i^{th}$ trading period

**Definition 3.1.11 (Transaction scale)** To reduce the transaction cost, we need to control the transaction scale at each trading period. We define $o_t$ to represent the transaction scale at the end of the $t-1^{th}$ trading period, i.e.,

$$o_t = |\Delta q_t|^T p_{t-1}$$

The ratio between the scale of stock trading and the total investment scale is defined as the transaction scale ratio $\epsilon_t$, i.e.,

$$\epsilon_t = \frac{|\Delta q_t|^T p_{t-1}}{T_t} \tag{8}$$

**Definition 3.1.12 (Price fluctuation tensor)** Considering that historical return distribution is essential information to the DRL agent, we formulate the price fluctuation tensor $Y_t \in \mathbb{R}^{1 \times 5 \times n}$ to describe the price fluctuation within each trading period:

$$Y_t = [z_{t_1}, z_{t_2}, z_{t_3}, z_{t_4}, z_{t_5}].$$

## 3.2 Trading process

We assume that there are no short-sales constraints in the stock market. As given in Figure 2, the portfolio weights $w_t$ are calculated at the end of each trading period. The corresponding target quantity $q_t$ can then be calculated according to Equation (4). If the target quantity for asset $i$: $q_{i,t} \geq 0$, investors should hold $q_{i,t}$ shares of stock $i$ in the long position at the $t^{th}$ trading period. If $q_t < 0$, the investor should hold $-q_{i,t}$ shares of stock $i$ in the short sale. To realize the target quantity $q_{i,t}$ at the $t^{th}$ trading period, investors should perform the following steps at the end of each trading period: (1) Calculate the market order vector $\Delta q_t$ according to the target quantity $q_t$ and the holding position $q_{t-1}$ in the last trading period. (2) For stock $i, \{i = 0, 1, 2, \ldots, n\}$ in the portfolio, investors should buy $\Delta q_{i,t}$ shares of stocks when $\Delta q_{i,t} \geq 0$. When $\Delta q_{i,t} < 0$, investors should establish a short position or close up $\Delta q_{i,t}$ shares of stock.

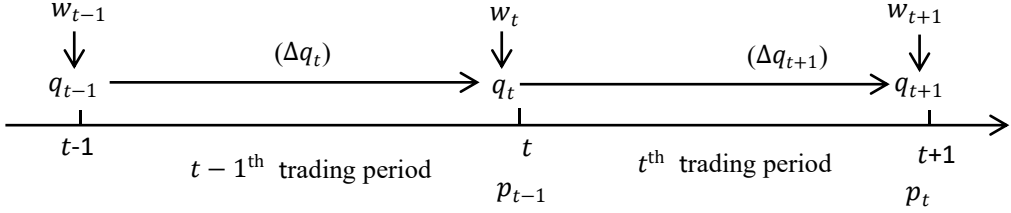

Fig 2. The trading process in each trading period. The target portfolio weights $w_t$ are determined by the DRL agent at the end of the $t - 1^{th}$ trading period. According to the target portfolio weights, the corresponding target quantity $q_t$ and market order $\Delta q_t$ can be calculated.

## 3.3 Problem Setting

Here, we shall define the periodical portfolio management problem. The dynamic decision process of periodical portfolio management fits naturally into the framework of the finite Markov Decision Process, which is defined by the tuple $M =< \mathcal{S}, \mathcal{A}, \mathcal{P}, \mathcal{R} >$. In the Markov Decision Process $M =< \mathcal{S}, \mathcal{A}, \mathcal{P}, \mathcal{R} >$, $\mathcal{S}$ represents the state space, and $\mathcal{A}$ represents the action space. The state and action space are continuous in the Markov Decision process. $\mathcal{P}: \mathcal{S} \times \mathcal{A} \rightarrow \mathcal{S}$ is the state transition dynamics, defined as a probability distribution $P(s_{t+1}|s_t, a_t)$, and $\mathcal{R}$ is the reward function. In the $t^{th}$ trading period, the DRL agent observes the state $s_t \in \mathcal{S}$ and makes the action $a_t \in \mathcal{A}$ based on its policy $a_t = \pi_\phi(s_t)$. Subsequently, the environment transits to the next state $s_{t+1}$ based on the transition model $\mathcal{P}$. According to state $s_{t+1}$, the DRL agent receives the reward $r_t$ from the reward function $\mathcal{R}$. In the portfolio management problem, the DRL agent determines the assets reallocation actions $\{a_1, a_2, \ldots, a_T\}$ periodically according to the current market environmental state $\{s_1, s_2, \ldots, s_T\}$ and acts on these portfolio weights at the end of each trading period. The agent's goal is to learn the policy that can maximize the return per unit of risk in consecutive trading periods.

## 3.4 Assumptions

In this research, our experiments are all based on back-test tradings in which the DRL agent is assumed to begin trading at a historical time point with no prior knowledge of the future. To meet the requirement of the back-test tradings set in each experiment, we make the following assumptions:

(1) Zero Market Impact: The trading actions made by the agent do not affect the asset price fluctuation.

(2) Zero Slippage: The selected assets in the portfolio have notably abundant liquidity in the market. Hence, whether the long position trade orders or short selling orders are issued, they can be immediately executed after they are issued.

In a real trading environment, these two assumptions are valid when the trading volume of the target asset is high.

## 4. Methodology

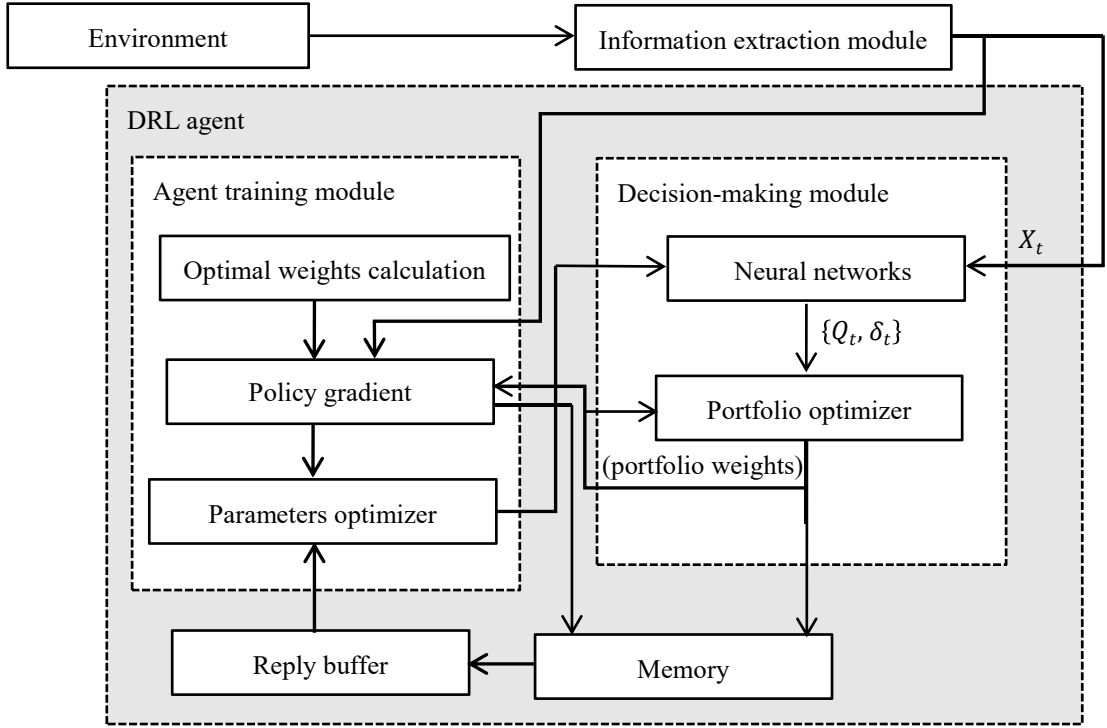

Fig 3. DRL agent framework. This framework comprises three modules: an information extraction module, an agent training module and a decision-making module. The information extraction module collects data from the environment and describes the environmental state $s_t$ of the DRL agent. When receiving the state $s_t$, the decision-making module determines the portfolio weights $w_t$. The decision-making module of the DRL agent consists of the neural networks and BL-based portfolio optimizer module. The agent training module calculates the theoretical optimal portfolio weights based on the state $s_{t+1}$ in the next trading period. According to the optimal portfolio weights, the agent training module calculates the policy gradient and updates the parameters in the policy function.

As aforementioned, we highlight the importance of learning the correlation between portfolio asset returns when applying the long/short strategy. Simultaneously, we also note that it is a challenge for DRL agents to learn the correlation of the portfolio asset returns and implement a long/short strategy based on the correlation. To address this, we propose a multi-module DRL trading system to train the DRL agent to apply the BL model for portfolio optimization.

The design of the framework is given in Figure 3. In this framework, the multi-module DRL trading system comprises an information extraction module, a decision-making module, and an agent training module. We will next describe the specific function of these three modules in detail.

## 4.1 Information extraction module

In the DRL framework, the information extraction module specifies the DRL agent's state. As given in Figure 3, the DRL agent retrieves the state description from the information extraction module. By observing the current state of the environment, the DRL agent determines the target portfolio weights and reallocates assets at the end of each trading period.

To describe the state $s_t$ at the end of the $t-1^{\text{th}}$ trading period, the information extraction module collects the price fluctuation tensor $Y_t \in \mathbb{R}^{1 \times 5m \times n}$ in the past $m$ trading periods to construct the historical return tensor:

$$X_t = [Y_{t-m}, Y_{t-m+1}, \dots, Y_{t-1}].$$

Since the historical return tensor $X_t$ reflects the historical distribution of the asset returns, the prior distribution parameters in the BL model can be calculated based on the price tensor $X_t$. Furthermore, the target portfolio weights $w_{t-1}$ in the previous trading period are also important information for the DRL agent because the difference between the portfolio weights $w_{t-1}$ and $w_t$ needs to be controlled to avoid high transaction costs. We define the state $s_t$ as a two-tuple:

$$s_t := \langle w_{t-1}, X_t \rangle$$

where $w_{t-1}$ is the portfolio weights determined at the end of the last $t-1^{\text{th}}$ trading period, and $X_t$ is the historical return tensor at the $t^{\text{th}}$ trading period as defined above.

## 4.2 Decision-making module

To enable the DRL agent to effectively learn the policy to utilize the BL model for portfolio decision-making, we design a novel set of decision rules for the DRL agent. Here, we shall describe in detail the design of the decision process of the target portfolio weights $w_t$ based on the current state $s_t$.

The DRL agent will reallocate assets based on the current state at the end of each trading period. Since the DRL agent determines sequential portfolio weights $\{w_1, w_2, \dots, w_T\}$ according to the DRL agent's state, the action $a_t$ of the agent is defined as the portfolio weights:

$$a_t := w_t.$$

The DRL agent makes decisions based on the BL model and its subjective views $Q_t$ for the return expectation $\mu_t$ during the current trading period. The BL model is a portfolio optimization model based on the Bayesian framework. It allows us to calculate the distribution of the excess return by combining the views of the expected excess return with the historical data. Since we assume that the risk-free return is zero, the risk assets return $z_{t_k}$ given in Definition 3.1.5 is equivalent to the excess return of the BL model. Hence, we shall refer to $z_{t_k}$ as the excess return vector. It is assumed that the excess return vector $z_{t_k}$ follows the conditional normal distribution [44]:

$$z_{t_k} | \mu_t \sim N(\mu_t, \Sigma_t).$$

The distribution of the expectation of excess return $\mu_t$ is the prior distribution in the Bayesian framework. The prior distribution of the expected excess return vector is defined as

$$\mu_t \sim N(\Pi_t, \tau \Sigma_t), \qquad (9)$$

where $\tau$ is a scalar describing the confidence level of the prior expectation $\Pi_t$. The covariance matrix

$\Sigma_t$ is estimated using historical return data:

$$\Sigma_t := \Sigma_t^h = \frac{\left[X_t - \frac{1}{5m}\sum_{i=1}^{m}\sum_{k=1}^{5} z_{t-i_k}\right]\left[X_t - \frac{1}{5m}\sum_{i=1}^{m}\sum_{k=1}^{5} z_{t-i_k}\right]^T}{5m - n - 1}, \quad (10)$$

where $n$ is the number of assets in the portfolio, and $m$ is the number of trading periods included in covariance estimation.

The expectation $\Pi_t$ of the prior distribution is the neutral starting point in the BL model [45]. In the search for a reasonable starting point for expected return, Black, Litterman, and He [46-48] explore several alternative forecasts: historical return, equal mean return for all assets, risk-adjusted equal mean return and the implied equilibrium excess return. In our framework, the equilibrium return is derived based on the assumption that each portfolio asset has an equivalent investment value, and such equilibrium return is the neutral starting point in the BL model. Hence, the prior expected return $\Pi_t$ is calculated by using the inverse optimization method [45]:

$$\Pi_t = \Sigma_t^h \frac{1}{n\delta_t} e, \quad (11)$$

where $e$ is the vector of ones, $\delta_t$ is the risk aversion of the DRL agent, $n$ is the number of assets in the portfolio, and $\Sigma_t^h$ is the covariance matrix calculated using Equation (10). The prior expected excess return $\Pi_t$ is the value of return expectation when the solution to the following unconstrained objective function maximization problem

$$\begin{cases} \max_{w_t} & w_t^T \Pi_t - \lambda_t w_t^T \Sigma_t^h w_t \\ \text{s.t.} & w_t^T e + w_{0,t} = 1 \end{cases},$$

is

$$w_t = \frac{1}{n} e.$$

Part of the innovation of the BL model is that subjective views $Q_t$ regarding the expectation of the excess return $\mu_t$ can now be adopted as the model's inputs. The views about the expectation of the assets' excess return are all expressed by linear equations, i.e., the $i^{\text{th}}$ view at the $t^{\text{th}}$ trading period $Q_{i,t}$ can be represented by

$$Q_{i,t}: \omega_{i1,t}\mu_{1,t} + \omega_{i2,t}\mu_{2,t} + \ldots + \omega_{in,t}\mu_{n,t} = Q_{i,t} + \varepsilon_{i,t}, \quad (12)$$

where $\omega_{ij,t}$ is the weight of asset $j$ in view $i$ at the $t^{\text{th}}$ trading period and $\mu_{j,t}$ denotes the expected return of asset $j$ at the $t^{\text{th}}$ trading period for $j = 1, 2, \ldots, n$. The linear equation reflects the view of the linear combination of the expected excess return of the assets. Equation (12) reflects the view that the expectation of the portfolio with portfolio weights $\omega_{i,t} = [\omega_{i1,t}, \omega_{i2,t}, \ldots, \omega_{in,t}]^T$ follows the normal distribution, where the mean is $Q_{i,t}$ and the variance is $\text{Var}(\varepsilon_{i,t})$. All views are mutually independent. There are two kinds of views in the BL model: absolute view [45] and relative view [45]. Since we train the DRL agent to provide return expectations for each portfolio asset, the views proposed by the DRL agent are adopted as the absolute views [45] in the BL model. For the absolute view [45], the sum of the weights for the assets is one. The framework for the stated views in the BL model can be expressed as:

$$P\mu_t = Q_t + \varepsilon_t,$$

where $P$ is a $v \times n$ fixed matrix identifying the assets involved in each view. $\mu_t$ is the expected excess return at the $t^{\text{th}}$ trading period. $Q_t$ is a $v \times 1$ vector representing the views of the linear combination of the portfolio assets' expected excess return at the $t^{th}$ trading period. $\varepsilon_t$ represents the uncertainty of the views, which follow the normal distribution:

$$\varepsilon_t \sim N(\mathbf{0}, \Omega_t),$$

where $\mathbf{0}$ is a $v \times 1$ zero vector, and $\Omega_t$ is a $v \times v$ diagonal covariance matrix such that

$$\Omega_t = diag(P(\tau \Sigma_t^h) P^T).$$

In the decision process, the DRL agent applies a neural network to provide its subjective views $Q_t$ for the excess return expectation $\mu_t$ at the end of each trading period. It then adopts the BL model to calculate the portfolio weights $w_t$ based on its subjective views $Q_t$ and risk aversion $\delta_t$. According to the BL model, the expectation of the excess return on the assets combining the prior return distribution described in Equation (9) and the DRL agent's subjective views $Q_t$ follow a normal distribution:

$$\mu_t | Q_t \sim N(\mu_t^{BL}, \Sigma_{\mu_t}^{BL}).$$

Based on the Bayesian framework [49], the combined posterior distribution of the excess return expectation $\mu_t$, which combines the prior excess return $\Pi_t$ and subjective views $Q_t$ of the DRL agent, can be expressed as:

$$\mu_t | Q_t \sim N\big([(\tau \Sigma_t^h)^{-1} + P^T \Omega_t^{-1} P]^{-1} [(\tau \Sigma_t^h)^{-1} \Pi_t + P^T \Omega_t^{-1} Q_t], [(\tau \Sigma_t^h)^{-1} + P^T \Omega_t^{-1} P]^{-1}\big).$$

Conditioned on the return expectation $Q_t$ of the DRL agent, the distribution of the excess return calculated by the BL model can be expressed as

$$z_{t_k} | Q_t \sim N\big([(\tau \Sigma_t^h)^{-1} + P^T \Omega_t^{-1} P]^{-1} [(\tau \Sigma_t^h)^{-1} \Pi_t + P^T \Omega_t^{-1} Q_t], \Sigma_t^h + [(\tau \Sigma_t^h)^{-1} + P^T \Omega_t^{-1} P]^{-1}\big). \quad (13)$$

The DRL agent then calculates the optimal portfolio weights based on the view that the excess return follows the normal distribution described in Equation (13), where expectation $\mu_t^V$ and variance $\Sigma_t^V$ can be expressed as:

$$\Sigma_t^V = \Sigma_t + [(\tau \Sigma_t^h)^{-1} + P^T \Omega_t^{-1} P]^{-1}, \text{ and}$$

$$\mu_t^V = [(\tau \Sigma_t^h)^{-1} + P^T \Omega_t^{-1} P]^{-1} [(\tau \Sigma_t^h)^{-1} \Pi_t + P^T \Omega_t^{-1} Q_t].$$

Subsequently, the DRL agent calculates its current risk aversion $\delta_t$ based on the state $s_t$. In the decision process of the DRL agent, the portfolio weights are obtained by solving the following concave quadratic programming problem [50]:

$$\begin{cases} \min_{w_t} \frac{1}{2} w_t^T \Sigma_t^V w_t - \delta_t w_t^T \mu_t^V \\ \text{s.t.} \quad w_t^T e + w_{0,t} = 1 \end{cases}, \quad (14)$$

where $e$ is the unit vector $[1,1,\ldots,1]^T$ and $\delta_t$ is the risk aversion parameter. According to the Lagrangian algorithm [50], this concave quadratic programming problem has a unique primal-dual solution:

$$\begin{cases} w_t = \delta_t \Sigma_t^{V^{-1}} \mu_t^V \\ w_{0,t} = 1 - w_t^T e \end{cases}, \quad (15)$$

where $w_t$ and $w_{0,t}$, respectively, represent the target weights of the risk assets and the risk-free asset at the beginning of the $t^{th}$ trading period. The portfolio management decision process is given in Algorithm 1.

**Algorithm 1. Portfolio management decision process of the DRL agent**

Input: The state $s_t$

1. Receive the state $s_t$ from the environment.

2. Neural network $N_1(s_t; \phi_1)$ is applied to simulate investor to output specific views $Q_t$ regarding the expectation of excess return $\mu_t$ of each asset in the portfolio:

$$Q_t = N_1(s_t; \phi_1),$$

where $\phi_1$ are the parameters of the neural network. Neural network $N_2(s_t; \phi_2)$ is applied to output the trade-off parameter $\delta_t$:

$$\delta_t = N_2(s_t; \phi_2),$$

which represents the risk aversion of the investment action.

3. $Q_t$ can be obtained as a $k \times 1$ vector, and it represents the DRL agent's subjective views of excess return on each portfolio asset. Since $Q_t$ is the expectation of the excess return of each asset in the portfolio, $k = n$. Hence, the corresponding view matrix $P$ is an $n \times n$ identity matrix, which reflects that all the views of the DRL agent are absolute views [45].

4. Obtain the prior distribution of return expectation $\mu_t$ based on the historical excess return data $X_t$ using Equation (10) and (11).

5. BL model is used to calculate the posterior distribution of return $z_{t_k}$ defined in Equation (13), which combines the subjective return expectation $Q_t$ and prior distribution described in Equation (9), (10), and (11).

6. Based on the posterior distribution of return $z_{t_k}$ described in Equation (13) calculated by the BL model and risk aversion $\delta_t$, construct the unconstrained portfolio optimization problem described in Equation (14).

7. Solve the portfolio optimization problem according to the Lagrangian algorithm [50] to obtain the optimal portfolio weights $w_t$ and $w_{0,t}$ using Equation (15).

8. Calculate the market order $\Delta q_t$ based on the calculated portfolio weights vector $w_t$ and reallocate the assets.

This section describes the decision process of the DRL agent. The detailed training process and the components of the DRL framework are described in the next section.

### 4.3 Agent training module

Since portfolio management can be formulated as a Markov decision process, the policy gradient method is adopted to optimize the investment policy in an off-policy manner. We adopt the policy-only method to train the DRL agent effectively and to avoid the gradient explosion in the training process. According to the decision-making process described in Algorithm 1, the deterministic policy function of the DRL agent can be summarized as follows:

$$\begin{cases} a_t := w_t = \pi_\phi(s_t) = \delta_t {\Sigma_t^V}^{-1} \mu_t^V \\ \Sigma_t^V = \Sigma_t^h + [(\tau \Sigma_t^h)^{-1} + P^T \Omega_t^{-1} P]^{-1} \\ \mu_t^V = [(\tau \Sigma_t^h)^{-1} + P^T \Omega_t^{-1} P]^{-1}[(\tau \Sigma_t^h)^{-1} \Pi_t + P^T \Omega_t^{-1} Q_t] \\ \Pi_t = \Sigma_t^h \dfrac{1}{n\delta_t} e \\ Q_t = N_1(s_t; \phi_1) \\ \delta_t = N_2(s_t; \phi_2) \end{cases}, \quad (16)$$

where the output of the policy function is defined as the portfolio weights $w_t$. When the asset reallocation is carried out based on the portfolio weights $w_t$ at the end of the $t-1^{\text{th}}$ trading period, the environment's reward function $\mathcal{R}$ evaluates the performance of the deterministic policy based on the state $s_{t+1}$ at the end of the $t^{\text{th}}$ trading period. The agent's goal, defined in Section 3.3, is to learn the policy that can maximize the return per unit of risk in consecutive trading periods. In our framework, the reward function $\mathcal{R}$ in the environment is constructed based on the daily return of the portfolio $w_t$, adjusted by the corresponding variance and the ratio of transaction scale, which is defined as:

$$\mathcal{R}: r^e(w_t|s_{t+1}, \lambda_1, \lambda_2) = \frac{1}{5}\xi_t - \frac{\lambda_1}{2}V_t - \frac{\lambda_2}{2}\epsilon_t, \tag{17}$$

where the term $\xi_t$ is the portfolio return in the $t^{\text{th}}$ trading period defined in Equation (6). Since there are five trading days in a trading period, the term $\frac{1}{5}\xi_t$ represents the average daily return of the portfolio in the $t^{\text{th}}$ trading period. Furthermore, in the reward function $\mathcal{R}$, the term $V_t$ is the variance of the portfolio defined in Equation (7), and the term $\epsilon_t$ is the transaction scale ratio defined in Equation (8). During consecutive trading periods, a large transaction scale could result in high transaction costs and increased uncertainty of the return. Hence, we need to constrain the variance and transaction scale simultaneously. The parameters $\lambda_1$ and $\lambda_2$ in Equation (17) are positive constants.

The training objective is to maximize the accumulated value of the reward (ARD$^{(\text{tr})}$) that our DRL agent obtains from the training environment:

$$\text{ARD}^{(\text{tr})} = \sum_{i=1}^{T_f^{(tr)}} r^e(w_t|s_{t+1}, \lambda_1, \lambda_2). \tag{18}$$

where $T_f^{(tr)}$ is the total number of trading periods in the training environment.

In traditional actor-critic algorithms, a critic network is trained to approximate the expectation of the reward received by the DRL agent after taking action $a_t$ in the state $s_t$ and thereafter following the policy $\pi_\theta$. The policy function of the DRL agent is updated by propagating the critic network's gradient to the policy function of the DRL agent. However, in the portfolio optimization problem, the action space is continuous and high dimensional. The problem of the curse of dimensionality [51] always exists when applying the actor-critic algorithm for DRL agent training. The curse of dimensionality refers to the exponential growth of states and actions when exploring optimal policy in high-dimensional spaces [51]. As a result, DRL agents suffer from poor sample efficiency and poor scalability [52], and the well-training of the critic network is challenging to realize.

To avoid this, we formulate a differentiable training objective function of the target portfolio weights $w_t$ according to the reward function $\mathcal{R}$ in the environment. In the formulation process of the training objective function, we first formulate an evaluation function $\rho(w_t|s_{t+1}, \lambda_1, \lambda_2)$ based on the reward function $\mathcal{R}$ in the environment:

$$\rho(w_t|s_{t+1}, \lambda_1, \lambda_2) = w_t^T \mu_t - \frac{\lambda_1}{2} w_t^T \Sigma_t w_t - \frac{\lambda_2}{2} |w_t^T - w_{t-1}^T|e, \tag{19}$$

In the evaluation function $\rho(w_t|s_{t+1}, \lambda_1, \lambda_2)$, $\mu_t$ is the expectation of the return vector defined in Equation (2). It is calculated based on the observed values of $z_{t_k}, k = 1,2,3,4,5$ in the $t^{\text{th}}$ trading period:

$$\mu_t = \frac{1}{5}\sum_{k=1}^{5} z_{t_k},$$

the term $w_t^T \mu_t$ is the estimation function of the return expectation, which is calculated based on the term $\frac{1}{5}\xi_t$ in the environment's reward function $\mathcal{R}$ defined in Equation (17). Consistent with the calculation of the variance measure in the reward function $\mathcal{R}$, the term $w_t^T \Sigma_t w_t$ is used to calculate the variance of the portfolio. The term $|w_t^T - w_{t-1}^T|e$ is used to estimate the transaction scale ratio $\epsilon_t$ in the reward function $\mathcal{R}$.

To avoid overfitting and realize the generalization ability of our BDA's policy, we do not choose to update our agent by maximizing the evaluation function. Instead, we set a target value $\Gamma_t$ for the evaluation function $\rho(w_t|s_{t+1}, \lambda_1, \lambda_2)$, and the goal of training the DRL agent is to let the evaluation function achieve this target value. In this paper, we determine target values $\Gamma_t$ by computing the theoretically optimal portfolio weights under the unconstrained portfolio optimization problem with the risk aversion $\lambda_3$:

$$w_t^{optimal} = \frac{1}{\lambda_3} \Sigma_t^{-1} \mu_t.$$

The evaluation function value of the theoretically optimal investment portfolio will serve as the target value for the DRL agent:

$$\Gamma_t = \rho(w_t^{optimal}|s_{t+1}, \lambda_1, \lambda_2) = w_t^{optimal\,T} \mu_t - \frac{\lambda_1}{2} w_t^{optimal\,T} \Sigma_t w_t^{optimal} - \frac{\lambda_2}{2} \left| w_t^{optimal\,T} - w_{t-1}^T \right| e. \quad (20)$$

Since the target value $\Gamma_t$ is applied to limit the value of the evaluation function $\rho(w_t|s_{t+1}, \lambda_1, \lambda_2)$ in the training process, the target value $\Gamma_t$ should lower than the maximum value of the evaluation function $\rho(w_t|s_{t+1}, \lambda_1, \lambda_2)$. Hence, the risk aversion $\lambda_3$ should be much larger than the risk aversion $\lambda_1$.

In the training process, the policy function of the DRL agent is updated to maximize its evaluation function value to reach the target value $\Gamma_t$. Hence, the training objective function $\Theta(w_t|s_{t+1}, \lambda_1, \lambda_2)$ is defined in terms of this difference between the evaluation function value of the optimal portfolio weights $w_t^{optimal}$ and the portfolio weights output by our DRL agent's policy function $\pi_\phi(S_t)$.:

$$\begin{cases} \Theta(w_t|s_{t+1}, \lambda_1, \lambda_2) = -[\Gamma_t - \rho(w_t|s_{t+1}, \lambda_1, \lambda_2)]^2 \\ \Gamma_t = \rho(w_t^{optimal}|s_{t+1}, \lambda_1, \lambda_2) \end{cases}.$$

In this way, the optimization objective of the DRL agent is to maximize the training objective function for given trajectories. Hence, the gradient direction of the policy function parameters $\phi$ update can be calculated as

$$\nabla J_{PG}(\phi) = \frac{1}{N} \sum_{i=1}^{N} \nabla_\phi \Theta(\pi_\phi(s_i)|s_{i+1}, \lambda_1, \lambda_2).$$

where $N$ is the sample size.

### *4.4 Convergence performance tracking in the training process*

To demonstrate that our training algorithm can realize the convergence of the objective function and effectively improve our BDA's ability to obtain higher accumulated return and accumulated reward in the training environment, we borrow the ideas from [86-90], and evaluate our DRL agent in the training environment and systematically collect the critical performance metrics to characterize their respective evolutionary trends throughout the training process. The numerical change trajectories of the following metrics are tracked in the training process.

In the training process, the training objective of the policy function is to maximize the sampled objective function value (OP), which is defined as:

$$\text{OP} = \frac{1}{N}\sum_{i=1}^{N} \Theta\left(\pi_\phi(s_i)|s_{i+1}, \lambda_1, \lambda_2\right) \tag{21}$$

To evaluate the convergence performance of our DRL algorithm, we track the numerical change trajectories of OP. Since the training objective is to let the value of the evaluation function $\rho(w_t|s_{t+1}, \lambda_1, \lambda_2)$ defined in Equation (19) converge to the preset target value $\Gamma_t$ calculated based on Equation (20), we also track the numerical change trajectories of the sampled evaluation function value (EF):

$$\text{EF} = \frac{1}{N}\sum_{i=1}^{N} \rho(\pi_\phi(s_i)|s_{i+1}, \lambda_1, \lambda_2) \tag{22}$$

Furthermore, we need to demonstrate that our training algorithm can effectively improve our BDA's ability to obtain higher accumulated return and accumulated reward in the training environment. Consequently, within this evaluation module, except for tracking the changes in the value of the sampled objective function value (OP) and the sampled evaluation function value (EF) in the training process, we evaluate our DRL agent in the training environment upon the completion of each training stage. In each evaluation, we let our DRL agent determine the target portfolio weights at the end of all the trading periods in the training environment. Then, we calculate the accumulative return (AR$^{(\text{tr})}$):

$$\text{AR}^{(\text{tr})} = \log_2\left(\frac{v_{T_f^{(tr)}}}{v_0}\right), \tag{23}$$

and ARD$^{(\text{tr})}$ based on Equation (18).

The details of the training process are given in Algorithm 2, and the hyper-parameters are described in detail in Appendix 1.

---

**Algorithm 2. The training process of the DRL agent**

Input: a policy function $\pi_\phi(S_t)$, learning rates $\alpha$, initial parameters $\phi^{(0)}$. minibatch size $N$, target step $M$, the number of trading periods $T_f^{(t)}$ in the training set, total step $S^{total} = 3e5$.

1. Initialize the Accumulated steps $S = 0$, trading period $t = 1$, current total asset $v_0 = T_1 = 1e8$, train = True.
2. Build the replay buffer

(Update the reply buffer)

3. If train do:
4.     Initialize the reply buffer.
5.     For $n = 1, 2, \ldots, M$ do
6.         Select action $a_t$ based on the policy function and the current state $s_t$:

$$a_t = \pi_\phi(s_t).$$

7. Enter into the next trading period $t = t + 1$.
8. Observe the state $s_{t+1}$ from the environment.
9. Calculate the reward $r_t$ that the DRL agent can obtain from the environment based on the reward function defined in Equation (17).
10. Store the tuple $\{s_t, a_t, r_t, s_{t+1}\}$ into the reply buffer
11. If $t = T_f^{(t)}$ do:
12. Reset the trading period: $t = 1$.

(Train the policy function of the DRL agent)

13. For $n = 1, 2, \ldots, \lfloor M/N \rfloor$ do:
14. Randomly sample a mini-batch of the states $\{s_i, a_i, r_i, s_{i+1}\}_{i=1}^N$ from the reply buffer.
15. Update the parameters of the policy function $\phi^{(n)}$ using sampled policy gradient:

$$\phi^{(n+1)} = \phi^{(n)} + \alpha \nabla J_{PG}(\phi^{(n)})$$

(Evaluate the performance of BDA in the training environment and track the numerical change trajectory of the critical indices)

16. Record the current OP and corresponding EF.
17. Evaluate our DRL agent in the training environment and record the $AR^{(tr)}$ and $ARD^{(tr)}$.

(Calculate the total steps and decide whether to finish the training)

18. Update the accumulated steps:

$$S = S + N.$$

19. If $S > S^{total}$ do:
    train = false

## 5. The Network topology in the DRL framework

In the topology of the neural network applied in the policy function, we adopt the Transformer [20] as the network backbone to simulate the investor to output subjective views of the excess return in the neural network $N_1(s_t; \phi_1)$. In the case of the neural network $N_2(s_t; \phi_2)$, the Convolutional Network (CNN) [53] is the backbone of the network. In this section, we shall describe the topology of the neural network in the policy function in detail.

In our DRL agent framework, the historical return tensor $X_t$ contained in the tuple $s_t$ for describing the state is adopted as the input of the neural networks $N_1(s_t; \phi_1)$ and $N_2(s_t; \phi_2)$. As given in Figure 4, since the number of the channel in the input tensor is one, for the Transformer network $N_1(s_t; \phi_1)$, the historical return tensor $X_t$ can be directly decomposed into a sequence of 1D patches:

$$X_t = [z_{(t-m)_1}, z_{(t-m)_2}, z_{(t-m)_3}, z_{(t-m)_4}, z_{(t-m)_5}, \ldots, z_{(t-1)_5}]$$
$$= [x_1, \quad x_2, \quad \ldots, \quad x_{5m}].$$

Hence, the size of the patches $x_t$ in the sequence is $n \times 1$.

The patch sequence is directly used as the patch embedding (PE) sequence in the Transformer [20], i.e.,

$$PE = [x_1, \quad x_2, \quad \ldots, \quad x_{5m}]$$

Analogous to the Vision Transformer [20], a learnable embedding $x_{Q-pre} \in \mathbb{R}^{n \times 1}$ is appended to the patch embedding sequence [54]:

$$z_0 = [x_{Q-pre}, \quad x_1, x_2, \ldots, x_{5m}], \tag{24}$$

and the learnable embedding $x_{Q-pre}$ [20] is denoted as $z_0^0$ in the patch embedding sequences. For $x_{Q-pre}$, its corresponding state is the output of the Transformer encoder $z_L^0$ [54]. The fully connected feed-forward network is used to $z_L^0$ to generate the investor's views $Q_t$ of the BL model [44]. As shown in Figure 4, the historical return tensor $X_t$ is composed of the return vectors $x_i$ in different trading days, and the return vectors $x_i$ are in chronological order in historical return tensor $X_t$. Different from the Vision Transformer [20], the positional embeddings of historical return vectors $x_i, (i = 1, 2, \ldots, m)$ are ignored since we concentrate on learning to extract non-linear correlations from multiple concurrent time series of portfolio assets. Hence, the position embedding [20] should not be added to the patch embedding sequence. The embedding sequence $z_0$ would serve as the input to the Transformer encoder module [54].

The Transformer encoder module [54] is composed of identical alternating layers, and each layer contains a self-attention (SA) block [54] and an MLP block. We define the input of each altering layer in the Transformer encoder module as $z_{\ell-1}$ $(\ell = 1, 2, \ldots, N)$.

The embedding sequence after layer normalization [57] $LN(z_{\ell-1})$ is the input of the Self-attention (SA) block [54]. For each SA block, the input sequence would respectively map to queries $q$, keys $k$, and values $v$ sequences [54] by learned linear projection:

$$\begin{cases} q_{\ell-1} = LN(z_{\ell-1})E_Q^{\ell-1} \\ k_{\ell-1} = LN(z_{\ell-1})E_K^{\ell-1} \\ v_{\ell-1} = LN(z_{\ell-1})E_V^{\ell-1}, \end{cases}$$

where $LN$ represents the Layer normalization [55], and the linear projections are parameters matrices:

$$E_q^{\ell-1} \in \mathbb{R}^{d \times L_{model}}, E_k^{\ell-1} \in \mathbb{R}^{d \times L_{model}}, E_v^{\ell-1} \in \mathbb{R}^{d \times L_{model}}.$$

Based on the queries $q$, keys $k$, and values $v$ sequences, the output of the attention function in the Self-attention block [54] is defined as

$$Attention_{\ell-1}(q_{\ell-1}, k_{\ell-1}, v_{\ell-1}) = Softmax\left(\frac{q_{\ell-1}k_{\ell-1}^T}{\sqrt{d}}\right)(v_{\ell-1}),$$

where $\sqrt{d}$ is the scaling factor. Furthermore, the residual connection [58, 59] is applied after each Self-attention (SA) block [54].

$$z'_\ell = Attention_{\ell-1} + z_{\ell-1}$$

where $z'_\ell$ is adopted as the output of the SA block.

The Self-attention block is followed by layer normalization [55]. The sequence $LN(z'_\ell)$ now becomes the input to the MLP block [20], which consists of two fully connected layers with a GELU activation function in between. Similar to the SA block, the residual connection [56] is applied after the MLP block [20]:

$$z_\ell = MLP(LN(z'_\ell)) + z'_\ell, \quad \ell = 1 \ldots L,$$

where $z_\ell$ denotes the outputs of the $\ell_{th}$ altering layers. After $L$ iterations, the sequence $z_L$ serves as the Transformer encoder [54] outputs. The treating process of the alternating layers in the Transformer encode module is summarized in Algorithm 3.

**Algorithm 3: Treating process of the Transformer encoder block**

Input: $z_0 = [x_{Q-pre}; \; x_1; \; x_2; \; ...; \; x_{5m}]$ (Equation (24)).

For $\ell$ from 1 to $L$:

$$z'_\ell = SA(LN(z_{\ell-1})) + z_{\ell-1}$$
$$z_\ell = MLP(LN(z'_\ell)) + z'_\ell$$

Outputs: $z_L$

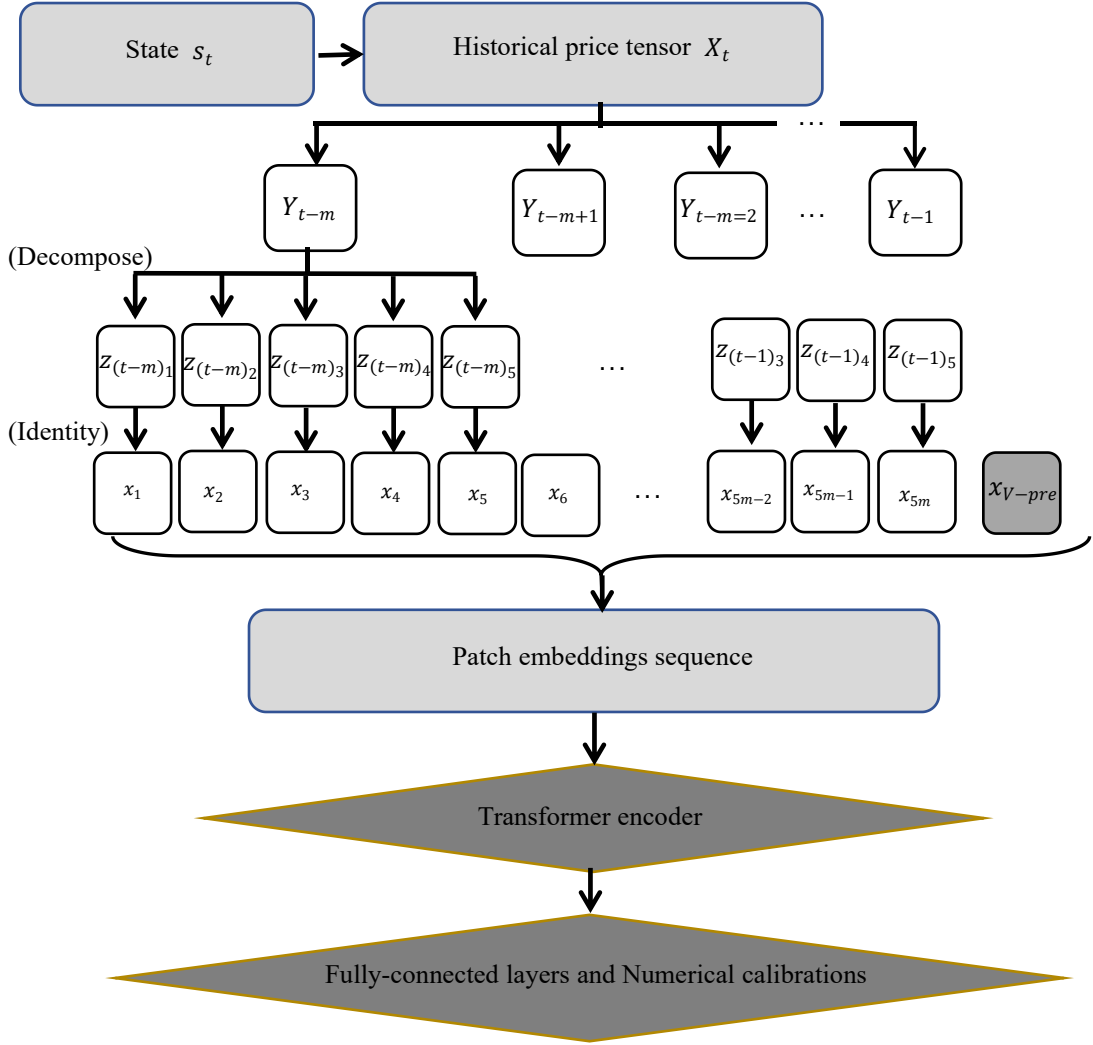

Fig 4. Transformer network topology. The historical price tensor in the state $s_t$ is the input of the transformer network. The historical return tensor $X_t$ is split into fixed patches in the time dimension. The resulting patch sequence is the patch embeddings, which is the input of the transformer encoder module. The output of the Transformer encoder is input into the fully connected layers and numerical calibrations module to get the outputs $Q_t$ of the neural network $N_1(s_t; \phi_1)$.

From the output of the Transformer encoder sequence $z_L$, we extract the $Q$-prediction token $z_L^0$, which is the first patch in sequence:

$$z_L = [z_L^0, \; z_L^1, \; ..., z_L^{5m}].$$

Since $Q$-prediction token $z_L^0$ is a 1-D tensor, it is fed into the fully connected feed-forward networks with a log-sigmoid activation function in between:
$$y = W_2(LogSigmoid(W_1 z_L^0))$$
where $W_1 \in \mathbb{R}^{n \times l}$ and $W_2 \in \mathbb{R}^{l \times n}$. The log-sigmoid activation function has a smooth gradient [57], which is beneficial for our gradient-based optimization methods because it allows for a more stable convergence. In addition, the saturation property [58] of the log-sigmoid function can realize regularization that helps mitigate the effects of outliers in the input data. For the predicted excess return to be more consistent with the return distribution, the output of the fully connected neural network is multiplied by the variance of each stock.
$$Q_t = diag(\Sigma_t)y.$$
Subsequently, the DRL agent can receive its subjective views $Q_t$ for the excess return. The detailed topology of the Transformer adopted in the neural network $N_1(s_t; \phi_1)$ is depicted in Figure 4.

To balance between the return and risk, we derive an appropriate risk aversion based on the current state $s_t$. We adopt the Convolutional Network (CNN) to output the risk aversion $\delta_t$. The detailed topology of the neural network $N_2(s_t; \phi_2)$ is depicted in Figure 5. After receiving the subjective views $Q_t$ for the excess return provided by the Transformer network $N_1(s_t; \phi_1)$ and the risk aversion $\delta_t$ provided by $N_2(s_t; \phi_2)$, the optimal portfolio weights can be directly calculated based on the policy function defined in Equation (16).

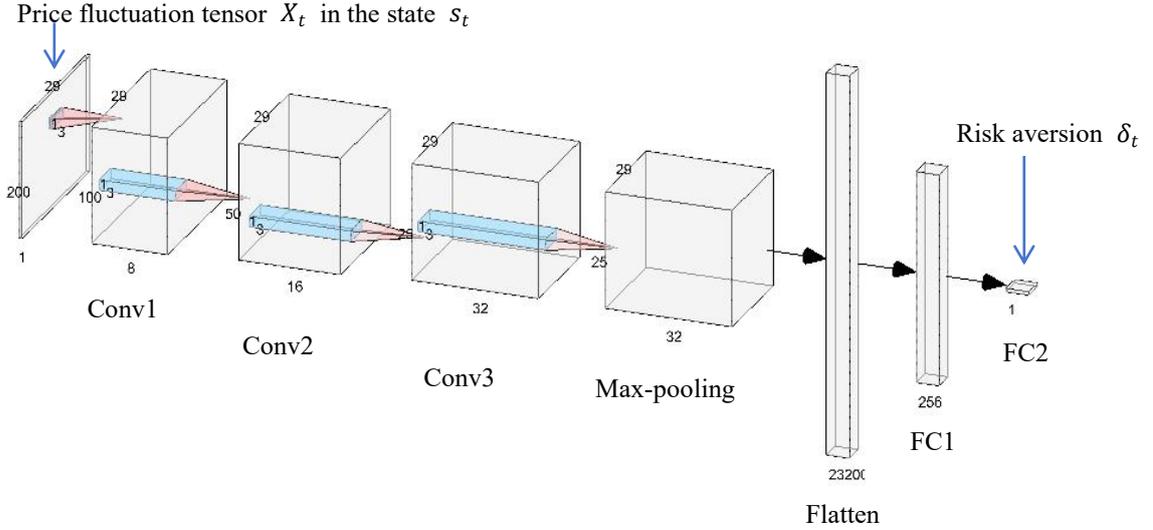

Fig 5. Topology of network $N_2(s_t; \phi_2)$. Network $N_2(s_t; \phi_2)$ is adopted to determine the risk aversion $\delta_t$. As shown in the figure, the neural network $N_2(s_t; \phi_2)$ is constructed by the convolutional layers (Conv), max-pooling layers and fully connected (FC) layers.

# 6. Empirical results

## 6.1 Data Description and Experiment Setting

The proposed BDA is tested on the American stock market. The Dow Jones Industrial Average (DJIA) is a stock market index of 30 prominent U.S. companies. The stock data are obtained

from[1]Yahoo Finance[1]. We use DJIA constituent stocks to construct our portfolio. After removing stocks with missing data, 29 constituent stocks remain. The main reason for using constituent stocks of the DJIA is to conform to the assumptions presented in Section 3.4, which call for the selected stocks to have substantial market liquidity. Hence, we select the leading stocks from various sectors to construct our portfolio. The constituent stocks of the DJIA perfectly meet these requirements. The information on portfolio stocks is described in detail in Appendix 2.

The time range of the stock data is from January 2018 to December 2022. As previously stated, BDA is tested using four different experiments. The time horizons of the training sets and back-test sets are given in Table 1. Based on the data in the training sets and back-test sets, we construct the training environment and back-test environment in each experiment. As mentioned, the risk-free rate for the risk-free asset (cash) is set as zero in each training environment and back-test environment. The lending rate is set according to the Federal Funds Rate. In the time horizon of all the experiments, the Federal Funds Rate varies between 0.06% and 4.33% annualized. For simplification, the lending rate in each experiment is uniformly set as the average value of the Federal Funds Rate. Hence, the annualized lending rate $r_l$ is set at 3%. Given that the stocks in the portfolio all have high liquidity, the lending rate of each stock $r_s$ in the portfolio is uniformly set at 3% annualized in all the experiments.

In the table, it can be observed that our DRL agent is updated every six months in consideration of the dynamic nature of financial markets [59]. According to the adaptive market hypothesis [59], market participants learn and adapt to changes in the market environment. As investors adjust their strategies, the policy of the deep reinforcement learning model may become obsolete, necessitating retraining to capture new patterns. The time horizon of the back-test set is aligned with the update cycle of the DRL agents. Furthermore, to ensure that the trading periods encompassed within the time horizon of the back-test set are all complete, we opt for a marginally early termination of the back-test. Thus, we set the back-test duration at 120 trading days. In accordance with the time horizon of the back-test sets, we determine that the time horizon for the training set is three years. This training set duration is instituted to ensure that the DRL agent can be exposed to a diversity of market conditions in the training process. which facilitates the agent's ability to adapt across varied market scenarios, enhancing the robustness of its policy. Meanwhile, we are concerned that further extension of the time horizon of the training set might lead the DRL agent to learn the outdated patterns that no longer hold [60]. We do not further extend the time horizon of the training set.

| Experiment | Training set | Back-test set |
| --- | --- | --- |
| 1 | 2018.01.01 - 2020.12.31 | 120 trading days starting from 2021.01.01 |
| 2 | 2018.07.01 - 2021.06.30 | 120 trading days starting from 2021.07.01 |
| 3 | 2019.01.01 - 2021.12.31 | 120 trading days starting from 2022.01.01 |
| 4 | 2019.07.01 - 2022.06.30 | 120 trading days starting from 2022.07.01 |

Table 1: The time horizons of training sets and back-test sets in different experiments. In each experiment, we construct the training environment based on the three-year stock data. After training our BDA in the constructed environment, we proceed to carry out the back-test experiment over the subsequent 120 trading days to assess the out-of-sample performance of our BDA.

---

[1] Http://www. finance.yahoo.com

## 6.2 In-sample performance in the training process

As previously mentioned, during the training process of our DRL agent, we track the numerical change trajectory of OP and the corresponding EF defined in Equations (21) and (22). Meanwhile, we track the numerical change trajectory of $AR^{(tr)}$ and $ARD^{(tr)}$ that our DRL agent can obtain from the training environment, which can be calculated based on Equation (18) and (23). The outcomes of this tracking are depicted in Figure 6.

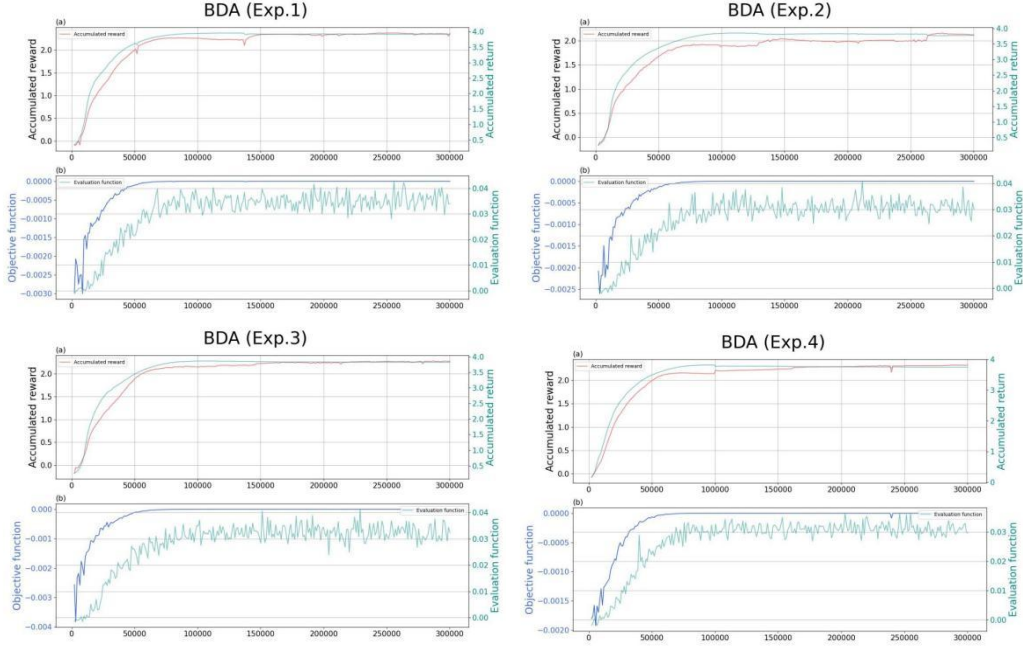

Fig 6. Track the numerical change trajectory of the objective function value (OP), evaluation function value (EF), accumulated return ($AR^{(tr)}$), and accumulated reward ($ARD^{(tr)}$) in the training process of BDA based on our policy gradient algorithm in four experiments given in Table 1. There are two subplots in each experiment's trajectory plot. Within the plot of each experiment, subplot (a) describes the numerical change trajectory of the accumulated return ($AR^{(tr)}$) and accumulated reward ($ARD^{(tr)}$) that our DRL agent can obtain from its training environment in the training process. Subplot (b) reflects the numerical change trajectory of the sampled objective function value (OP) and the sampled evaluation function value (EF) in the training process.

From the observed trajectories of OP and the corresponding EF in Figure 6, the training objective function and the evaluation function exhibit notable increments and ultimately achieve stable convergence in the training process. Notably, OP eventually converges towards zero, suggesting that the post-convergence fluctuation observed in EF is due to the numerical difference in the training target values $\Gamma_t$ across different trading periods. It indicates that the current policy gradient training algorithm founded upon the objective function $\Theta(w_t|s_{t+1}, \lambda_1, \lambda_2)$ can avoid the issues of gradient vanishing and gradient explosion during the training process. This training algorithm enables the evaluation function $\rho(w_t|s_{t+1}, \lambda_1, \lambda_2)$ value of our BDA's target portfolio weights $w_t$ to stably converge in the predetermined target value $\Gamma_t$.

As given in Figure 6, the numerical change trajectories of $AR^{(tr)}$ and $ARD^{(tr)}$ in four different experiments indicate that as the evaluation function steadily converges towards its target value, the values of the accumulated return and accumulated reward obtained by our BDA within the training environment exhibit a significant increase. Since the reward function $\mathcal{R}$ in the environment is constructed based on the daily return of the portfolio $w_t$ adjusted by the corresponding variance and the ratio of transaction scale, Its numerical change trajectory in our training process suggests that our policy gradient algorithm can significantly improve the profitability of our BDA, while simultaneously enhancing the agent's ability in controlling risk and trade scale in the training set.

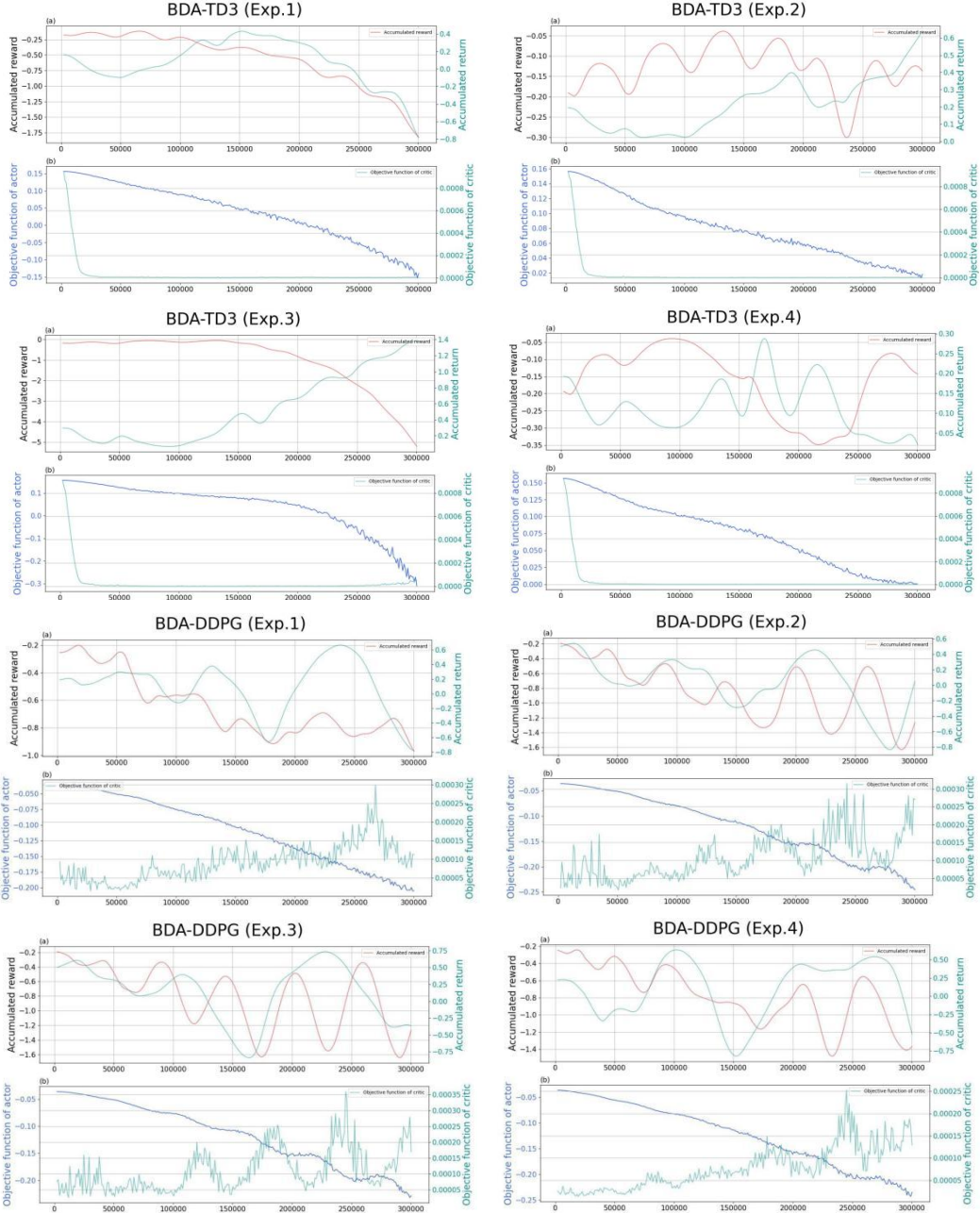

Fig 7. Track the numerical change trajectory of the sampled objective function value (OP), sampled evaluation function value (EF), accumulated return (AR$^{(tr)}$), and accumulated reward (ARD$^{(tr)}$) in the training process of the variants of BDA. In the variants, we adopt the DDPG algorithm and TD3 algorithm to replace our PG algorithm to train our BDA, respectively. The metrics displayed in the trajectory graphs of each experiment are consistent with those shown in Figure 6.

To substantiate the superiority of the current training methodologies over traditional deterministic policy actor-critic algorithms for the effective training of our agents, we implemented the Deep Deterministic Policy Gradient (DDPG) [33] algorithm and the twin delayed deep deterministic policy gradient (TD3) [35] algorithm as comparative benchmarks for training our BDA. Correspondingly, for the BDAs trained by the DDPG and TD3 algorithms (denoted as BDA-DDPG and BDA-TD3, respectively), I likewise track the numerical change trajectory of OP, EF, $AR^{(tr)}$, and $ARD^{(tr)}$. The numerical change trajectories of these metrics in each experiment are depicted in Figure 7.

From the observed numerical change trajectory of $AR^{(tr)}$ and $ARD^{(tr)}$ in Figure 7, we can find that training methodologies based on DDPG and TD3 algorithms do not improve the agent's capability to achieve higher $AR^{(tr)}$ and $ARD^{(tr)}$ in the training environment. This is due to the agent's continuous and highly dimensional action space. The training of the critic network faces the issue of the curse of dimensionality. Hence, the well-training of the critic network is difficult, and the gradient propagated from the critic network to the policy function of the DRL agent cannot improve its ability to obtain higher $AR^{(tr)}$ and $ARD^{(tr)}$ from the training environment. This suggests that, for the portfolio optimization problem, deriving the training objective function directly from the environment's reward function is more efficient than learning the value of the reward function via a critic network when the action space in the Markov Decision Process $M$ is continuous and high-dimensional.

## 6.3 Comparative strategies in the back-tests

To demonstrate that our BDA has outstanding performance in accumulated return and achieves a high level of return per unit of risk, the proposed BDA is compared to several well-known portfolio choice strategies in different back-test experiments. We divide the comparative strategies into three different categories as follows:

- **Traditional financial strategies:** Here, traditional financial strategies include strategies based on different capital growth theories and strategies based on Markowitz's [4] mean-variance theory.
- **Deep learning strategies:** In recent years, deep learning (DL) models have made considerable progress in return prediction. Researchers utilize the predictive returns output by these models to formulate investment portfolios and assess the robustness of these DL models by comparing their investment portfolios' performance in the back-test sets. Following the method applied by Duan et al. [61], here we implement TopK-Drop strategies[2] to determine the portfolio weights based on the prediction of the DL models. According to this strategy, within each trading period, investment capital is uniformly distributed among the top $K$ highest predicted return stocks.
- **Deep reinforcement learning strategies:** In this research, we use different portfolio optimization DRL algorithms as comparative strategies, i.e., DDPG [32], PPO [34], SAC [28], A2C [62], TD3 [57], and Policy Gradient (PG) [24]. In these DRL-based comparative strategies, DDPG [32], PPO [34], SAC [28], A2C [62], and TD3 [57] are adopted form the FinRL [32] platform. For the PG algorithm, we reproduce the DRL algorithm proposed by Jiang et al. [23].

The strategies of the different classes are described in detail in Table 2.

---

[2] https://qlib.readthedocs.io/en/latest/component/strategy.html

| | Traditional financial strategies | |
|---|---|---|
| Categories | Classifications | Algorithm |
| Strategies based on Capital Growth Theory | Baseline strategies | Constant Rebalanced Portfolios (CRP) [64] |
| | | M0 (M0) [65] |
| | | Uniform Buy And Hold (UBAH) [66] |
| | Follow-the-Winner | Universal Portfolio (UP) [64, 67] |
| | | Exponentiated Gradient (EG) [68] |
| | Follow-the-Loser | Anti-Correlation (ANTICOR) [69] |
| | | Passive Aggressive Mean Reversion (PAMR) [70] |
| | | Confidence Weights Mean Reversion (CWMR) [71] |
| | | Online Portfolio Selection with Moving Average Reversion (OLMAR) [72] |
| | | Robust Median Reversion (RMR) [73] |
| | | Weighted Moving Average Mean Reversion (WMAMR) [74] |
| | Pattern-Matching Approaches | Nonparametric Kernel Based Log Optimal Strategy (BK) [75] |
| | | Correlation-driven Nonparametric learning (CORN) [76] |
| | Meta-Learning Algorithm | Online Newton Step (ONS) [77] |
| Strategies based on Mean-Variance Theory | | Jorion's Bayes Stein procedure (JB) [78] |
| | | Kan and Zhou's three-fund rule (KZTF) [79] |
| | Strategies based on Machine learning methods | |
| Categories | | Algorithm |
| Strategies based on DL algorithms | | Dlinear [80] |
| | | Autoformer [61] |
| Strategies based on DRL algorithms | | EIIE [23] |
| | | DDPG-FinRL [23] |
| | | A2C-FinRL [23] |
| | | PPO-FinRL [23] |
| | | SAC-FinRL [23] |
| | | TD3-FinRL [23] |

Table 2: Comparative strategies. In the table, we detail the comparative strategies adopted in each back-test experiment.

## 6.4 Performance measure in the back-test experiments

Here, we adopt different performance metrics to evaluate the performance of our strategy in terms of profitability and risk. The transaction cost TC is set to 0.05% in the back-test experiments.

We note that the method of calculating returns in back-test experiments differs from the reward function calculation method used during the training of BDA. When calculating the reward function for the agent, logarithmic returns are computed based on the portfolio value change for each trading day.

To facilitate the comparison of different strategies in the back-test experiments, the logarithmic return of the investment at each trading day is calculated based on the change in the value of total assets. Since the investment amount is fixed as $0.5T_0$ at each trading period, the value of the portfolio is different from the value of total assets. The logarithmic return of the total assets $\vartheta_{t_k}$ at the $k^{\text{th}}$ trading day within the $t^{\text{th}}$ trading period in the back-tests is depicted as:

$$\vartheta_{t_k} = \begin{cases} \log_2(v_{t_k}/v_{t_{k-1}}) & k = 2,3,4,5 \\ \log_2(v_{t_k}/v_{t-1}) & k = 1 \end{cases} \quad (25)$$

In the back-test experiments, the adopted metrics include the accumulated return (AR), daily return (DR), standard deviation (Std), low partial standard deviation (LStd), Sharpe ratio (SR), and Sortino ratio (STR) of different strategies in the back-test experiments. The specific definitions of these metrics are described as follows.

Accumulated return (AR) is adopted to measure the profitability of the strategies. It is formally defined as

$$\text{AR} = \sum_{t=1}^{T_f} \sum_{k=1}^{5} \vartheta_{t_k}, \quad (26)$$

where $T_f$ denotes the number of trading periods in the back-tests, and $\vartheta_{t_k}$ is the logarithmic return of the total assets defined in Equation (25).

Daily return (DR) is the daily average of the logarithmic return $\vartheta_{t_k}$. It is calculated as

$$\text{DR} = \frac{1}{5T_f} \sum_{t=1}^{T_f} \sum_{k=1}^{5} \vartheta_{t_k}. \quad (27)$$

where $T_f$ and $\vartheta_{t_k}$ are as defined in Equation (26). Similar to AR, DR reflects the profitability of the strategies.

Variance (Var) and Standard deviation (Std) measure the risk of the strategies.

$$\text{Var} = \frac{1}{5T_f} \sum_{t=1}^{T_f} \sum_{k=1}^{5} (\vartheta_{t_k} - \text{DR})^2,$$

$$\text{Std} = \left[ \frac{1}{5T_f} \sum_{t=1}^{T_f} \sum_{k=1}^{5} (\vartheta_{t_k} - \text{DR})^2 \right]^{\frac{1}{2}}, \quad (28)$$

where DR is the daily return defined in Equation (27). $\vartheta_{t_k}$ and $T_f$ are as defined in Equation (26).

Sharpe ratio (SR) is a risk adjective return based on Daily Return (DR) and Standard deviation (Std). It represents the return on taking per unit of risk. It is formally defined as:

$$\text{SR} = \frac{\frac{1}{5T_f} \sum_{t=1}^{T_f} \sum_{k=1}^{5} \vartheta_{t_k} - r_f^d}{\sqrt{\frac{1}{5T_f} \sum_{t=1}^{T_f} \sum_{k=1}^{5} (\vartheta_{t_k} - \text{DR})^2}}, \quad (29)$$

where $r_f^d$ is the risk-free daily rate. Since the risk-free rate in the DRL agent's decision process is assumed to be zero, to maintain consistency, the risk-free daily rate in the performance measure is also assumed to be zero: $r_f^d = 0$. In Equation (29), the denominator is the Std defined in Equation (28), and $\vartheta_{t_k}$ and $T_f$ are as defined in Equation (26).

Some researchers [81] believe that the volatility caused by the positive return could not be viewed as a risk. Hence, we calculate the Lower partial Standard deviation (LStd):

$$\text{LStd} = \sqrt{\frac{1}{5T_f}\sum_{t=1}^{T_f}\sum_{k=1}^{5}(Min(\vartheta_{t_k}, c) - c)^2}, \tag{30}$$

where $c$ is the minimum acceptable return. The same as the risk-free rate $r_f^d$, we set the minimum acceptable return $c$ to zero.

Sortino ratio (STR) [81] is adopted to represent the gain on assuming per unit of downside volatility:

$$\text{STR} = \frac{\frac{1}{5T_f}\sum_{t=1}^{T_f}\sum_{k=1}^{5}\vartheta_{t_k} - c}{\sqrt{\frac{1}{5T_f}\sum_{t=1}^{T_f}\sum_{k=1}^{5}(Min(\vartheta_{t_k}, c) - c)^2}},$$

where the denominator is the low partial standard deviation (LStd) defined in Equation (30), and $c$ and $\vartheta_{t_k}$ are as defined in Equation (30).

## 6.5 Out-of-Sample Performance

The quantitative results of BDA and different portfolio choice comparison strategies in different back-test experiments are reported in Tables 3 - 6, respectively. In order to visualize the performance of BDA and other comparative strategies across various back-test experiments, the empirical results of various experiments are depicted in Figures 8 - 15. In Figures 8 - 15, four subplots are presented in each figure to visualize the empirical results. The four subplots visualize all numerical results from the back-test experiments documented in Table 3 - 6. It allows for an intuitive comparison of quantitative results across various strategies within the back-test experiments.

Specifically, subplot (a) of Figure 8 - 15 gives the AR trajectory of different strategies in four back-test experiments. Subplot (b) is a histogram reflecting the AR of different strategies in different time horizons of each back-test experiment. The empirical results reflected in subplots (a) and (b) show that our BDA achieves the highest accumulated return in all four experiments. Furthermore, BDA has advantages in AR over the comparison strategies in most of the back-test period in all four experiments. It demonstrates that BDA outperforms the comparative strategies by large margins in profitability.

The points in subplot (c) (Figures 8 - 15) represent different portfolio choice strategies. Each point's coordinates on the x-axis and y-axis represent the values of DR and Std that the strategies achieve in the back-test experiment, respectively. Hence, the slope of the line can reflect each strategy's SR. Similar to subplot (c), the coordinate value in the *y*-axis in subplot (d) also represents the DR. The coordinate value in the *x*-axis in subplot (d) represents the LStd. Hence, We can directly compare the STR of different strategies based on the slope of the line in subplot (d). The empirical results reflected in subplots (c) and (d) show that our BDA achieves the highest SR and STR in experiments 2, 3, and 4. Although our BDA does not achieve the highest SR and SRT in experiment 1, there is only a small gap between our BDA and the best-performing strategies (ONS). Moreover, the SR and STR achieved by our DRL agent are larger than 0.08 in all experiments. No comparative strategies can achieve this level. This suggests that our BDA can obtain outstanding performance in return per unit of risk by utilizing the dynamic correlation between the returns of portfolio assets and implementing the long/short strategy.

| Strategies | Performance Metrics | | | | | |
|---|---|---|---|---|---|---|
| | AR | DR | Std | SR | LStd | STR |
| BDA | **0.397118421** | **0.00330932** | 0.023735829 | 0.139422985 | 0.02380215 | 0.139034506 |
| BK | 0.018709122 | 0.000155909 | 0.01980568 | 0.007871951 | 0.022079669 | 0.007061218 |
| CRP | 0.199855819 | 0.001665465 | 0.010916012 | 0.152570838 | 0.010923113 | 0.152471653 |
| ONS | 0.207869288 | 0.001732244 | 0.010664163 | **0.162436007** | 0.010420531 | **0.166233764** |
| OLMAR | -0.059382698 | -0.000494856 | 0.02399772 | -0.020620951 | 0.024914275 | -0.01986234 |
| UP | 0.200112129 | 0.001667601 | 0.01092083 | 0.152699112 | 0.010928925 | 0.15258601 |
| Anticor | 0.173592878 | 0.001446607 | 0.01323422 | 0.109308092 | 0.013534095 | 0.106886148 |
| PAMR | -0.227382399 | -0.001894853 | 0.02250363 | -0.084202119 | 0.023740791 | -0.079814245 |
| CORNK | -0.066560018 | -0.000554667 | 0.021953059 | -0.025266038 | 0.024684278 | -0.02247045 |
| M0 | 0.186210139 | 0.001551751 | 0.012497954 | 0.124160412 | 0.013210835 | 0.117460494 |
| RMR | 0.073319683 | 0.000610997 | 0.02407251 | 0.025381539 | 0.025310865 | 0.024139727 |
| CWMR | -0.256554725 | -0.002137956 | 0.023229872 | -0.092034776 | 0.024416722 | -0.087561141 |
| EG | 0.185674288 | 0.001547286 | 0.011318425 | 0.136705034 | 0.01176915 | 0.131469625 |
| UBAH | 0.198802679 | 0.001656689 | 0.01195991 | 0.138520191 | 0.012301561 | 0.13467307 |
| WMAMR | -0.072117295 | -0.000600977 | 0.020129278 | -0.029855887 | 0.018293184 | -0.032852535 |
| JB | -0.100371303 | -0.000836428 | 0.004035638 | -0.207260309 | 0.004441518 | -0.1883202 |
| KZTF | -0.098616276 | -0.000821802 | 0.003915429 | -0.209888169 | 0.004429814 | -0.185516191 |
| FinRL-DDPG | 0.187725714 | 0.001564381 | 0.011793417 | 0.132648662 | 0.011777578 | 0.132827053 |
| FinRL-A2C | 0.066760423 | 0.000556337 | 0.004813015 | 0.11559009 | 0.004207106 | 0.132237428 |
| FinRL-PPO | 0.0629151 | 0.000524293 | 0.004595981 | 0.114076308 | 0.004099207 | 0.127900962 |
| FinRL-SAC | 0.197509898 | 0.001645916 | 0.01087987 | 0.151280831 | 0.010254317 | 0.160509551 |
| FinRL-TD3 | 0.189875871 | 0.001582299 | 0.015080461 | 0.104923777 | 0.014996067 | 0.105514261 |
| EIIE | 0.273840605 | 0.002282005 | 0.020908048 | 0.109144816 | 0.019566886 | 0.11662587 |
| Dlinear | 0.269446685 | 0.002245389 | 0.014128964 | 0.158920998 | 0.013581182 | 0.165330898 |
| Autoformer | 0.186644407 | 0.00155537 | 0.015804785 | 0.098411341 | 0.015865586 | 0.098034205 |

Table 3: Empirical results in experiment 1. The table depicts the quantitative results in accumulated return (AR), daily return (DR), standard deviation (Std), Sharpe ratio (SR), low partial standard deviation (LStd), and Sortino ratio (STR) of our BDA and various compared strategies in the back-test experiment 1. Transaction costs and lending costs are included in buying and selling actions. The best results for the metric regarding the return and return per unit of risk are highlighted in bold.

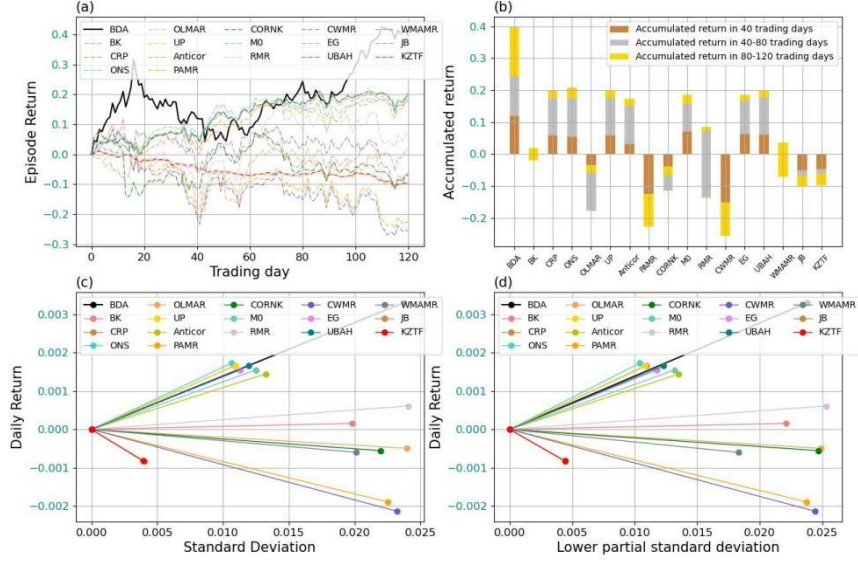

Fig 8. The visualization of the empirical results in Experiment 1. The figure reflects the empirical results of our BDA and the traditional comparative strategies.

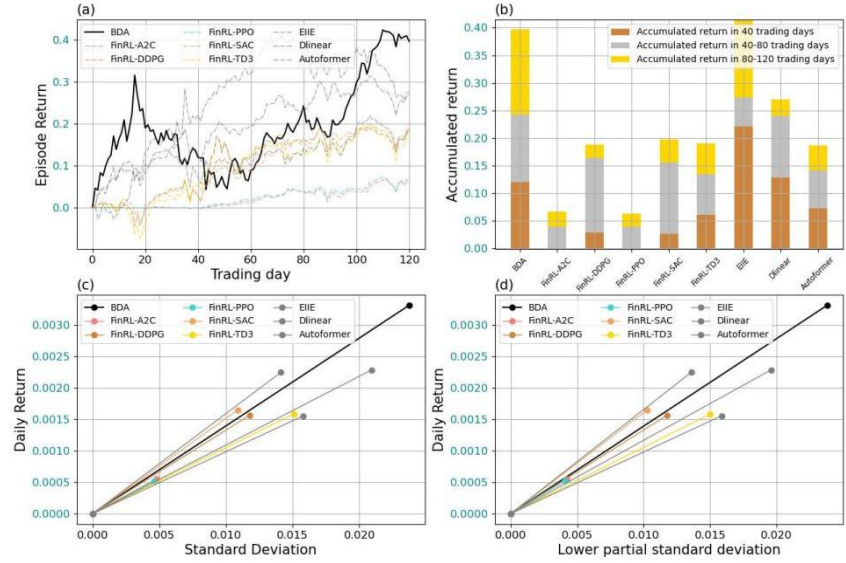

Fig 9. The visualization of the empirical results in Experiment 1. The figure reflects the empirical results of our BDA and the comparative strategies based on DL algorithms and DRL algorithms.

| Strategies | Performance Metrics | | | | | |
| --- | --- | --- | --- | --- | --- | --- |
| | AR | DR | Std | SR | LStd | STR |
| **BDA** | **0.274821202** | **0.002290177** | 0.026187598 | **0.087452722** | 0.024067521 | **0.09515632** |
| BK | -0.064098991 | -0.000534158 | 0.013891491 | -0.038452191 | 0.015889051 | -0.033618009 |
| CRP | 0.043105443 | 0.000359212 | 0.011011963 | 0.032620162 | 0.011557902 | 0.031079346 |
| ONS | 0.048112866 | 0.000400941 | 0.011272024 | 0.035569527 | 0.01174757 | 0.034129657 |
| OLMAR | -0.142524368 | -0.001187703 | 0.022379355 | -0.053071371 | 0.020951961 | -0.056686963 |
| UP | 0.043167695 | 0.000359731 | 0.011009991 | 0.032673124 | 0.011557193 | 0.031126138 |
| Anticor | 0.098244235 | 0.000818702 | 0.013703126 | 0.059745633 | 0.011935877 | 0.06859169 |
| PAMR | -0.247320985 | -0.002061008 | 0.02111851 | -0.097592499 | 0.022324886 | -0.092318869 |
| CORNK | -0.033457525 | -0.000278813 | 0.018259715 | -0.01526928 | 0.020657602 | -0.013496857 |
| M0 | 0.022049368 | 0.000183745 | 0.012537221 | 0.014655939 | 0.013400801 | 0.013711474 |
| RMR | -0.032826053 | -0.00027355 | 0.022823619 | -0.011985411 | 0.022053249 | -0.012404088 |
| CWMR | -0.257193108 | -0.002143276 | 0.021780742 | -0.098402337 | 0.022795989 | -0.094019867 |
| EG | 0.036989913 | 0.000308249 | 0.010845741 | 0.028421227 | 0.012058312 | 0.02556322 |
| UBAH | 0.035635485 | 0.000296962 | 0.010999321 | 0.026998247 | 0.011689147 | 0.025404965 |
| WMAMR | -0.056239114 | -0.000468659 | 0.021794947 | -0.021503117 | 0.020653007 | -0.022692061 |
| JB | 0.033133658 | 0.000276114 | 0.008549225 | 0.03229694 | 0.009627181 | 0.028680651 |
| KZTF | 0.04473883 | 0.000372824 | 0.007568315 | 0.049261109 | 0.007977728 | 0.04673305 |
| FinRL-DDPG | 0.0444739 | 0.000370616 | 0.011433985 | 0.032413532 | 0.012760722 | 0.029043484 |
| FinRL-A2C | 0.054652709 | 0.000455439 | 0.00587773 | 0.077485567 | 0.005638365 | 0.080775049 |
| FinRL-PPO | 0.007362327 | 6.13527E-05 | 0.006554506 | 0.009360388 | 0.00683896 | 0.008971061 |
| FinRL-SAC | 0.046983819 | 0.000391532 | 0.011947302 | 0.032771568 | 0.012591933 | 0.031093862 |
| FinRL-TD3 | -0.012637706 | -0.000105314 | 0.012179899 | -0.008646559 | 0.012433033 | -0.008470517 |
| EIIE | 0.053186513 | 0.000443221 | 0.014224683 | 0.031158581 | 0.01386439 | 0.031968297 |
| Dlinear | -0.061675786 | -0.000513965 | 0.013904575 | -0.036963725 | 0.014876659 | -0.034548409 |
| Autoformer | -0.037962934 | -0.000316358 | 0.01382982 | -0.022875047 | 0.013513026 | -0.023411322 |

Table 4: Empirical results in experiment 2. The metrics in this table are consistent with those in Table 3. Transaction costs and lending costs are included in buying and selling actions. The best results for the metric regarding the return and return per unit of risk are highlighted in bold.

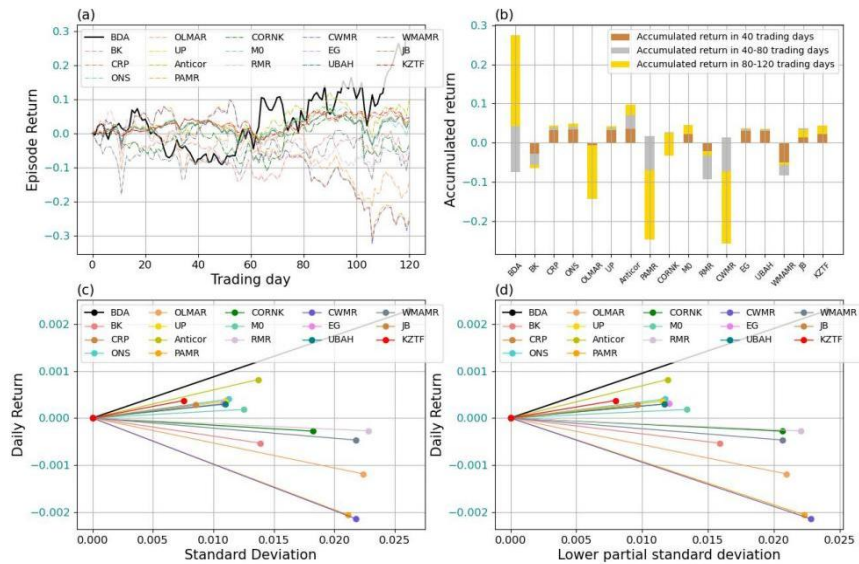

Fig 10. The visualization of the empirical results in Experiment 2. The figure reflects the empirical results of our BDA and the traditional comparative strategies.

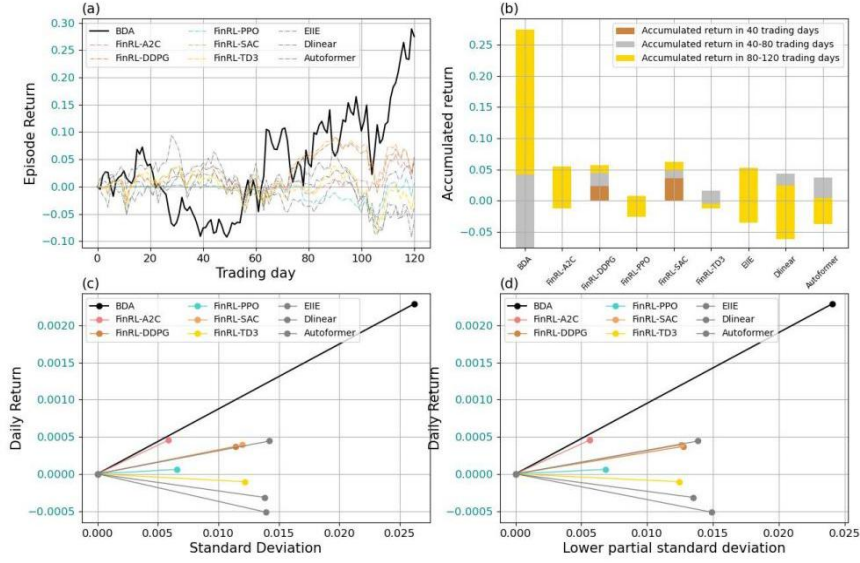

Fig 11. The visualization of the empirical results in Experiment 2. The figure reflects the empirical results of our BDA and the comparative strategies based on DL algorithms and DRL algorithms.

| Strategies | Performance Metrics | | | | | |
|---|---|---|---|---|---|---|
| | AR | DR | Std | SR | LStd | STR |
| BDA | **0.344678392** | **0.00287232** | 0.034691237 | **0.082796701** | 0.030119603 | **0.095363804** |
| BK | -0.128184848 | -0.001068207 | 0.022547292 | -0.047376291 | 0.026338033 | -0.040557587 |
| CRP | -0.184704273 | -0.001539202 | 0.018042228 | -0.085311098 | 0.019108399 | -0.080551085 |
| ONS | -0.192843436 | -0.001607029 | 0.019166901 | -0.083843947 | 0.019723264 | -0.081478838 |
| OLMAR | -0.505764117 | -0.004214701 | 0.034392127 | -0.122548423 | 0.036079856 | -0.116815904 |
| UP | -0.185251938 | -0.001543766 | 0.018043808 | -0.085556563 | 0.019116385 | -0.080756175 |
| Anticor | -0.165479562 | -0.001378996 | 0.025413382 | -0.054262606 | 0.025721709 | -0.053612158 |
| PAMR | -0.615434535 | -0.005128621 | 0.034780012 | -0.147458864 | 0.038222736 | -0.134177237 |
| CORNK | -0.151127458 | -0.001259395 | 0.023832573 | -0.052843455 | 0.026919394 | -0.046783947 |
| M0 | -0.236532138 | -0.001971101 | 0.019657423 | -0.100272611 | 0.021194639 | -0.092999989 |
| RMR | -0.678908415 | -0.00565757 | 0.03469974 | -0.16304359 | 0.03771874 | -0.149993615 |
| CWMR | -0.626530472 | -0.005221087 | 0.03574868 | -0.14604979 | 0.038794751 | -0.134582312 |
| EG | -0.17363428 | -0.001446952 | 0.016692104 | -0.086684837 | 0.017715475 | -0.081677311 |
| UBAH | -0.179313148 | -0.001494276 | 0.016100577 | -0.092808862 | 0.017283864 | -0.086454989 |
| WMAMR | -0.192366376 | -0.001603053 | 0.03294471 | -0.048658894 | 0.034792658 | -0.046074466 |
| JB | -0.129437256 | -0.001078644 | 0.009644273 | -0.111842932 | 0.01117308 | -0.096539524 |
| KZTF | -0.096954845 | -0.000807957 | 0.008452392 | -0.09558916 | 0.009650988 | -0.083717551 |
| FinRL-DDPG | -0.278260407 | -0.002318837 | 0.021038918 | -0.110216541 | 0.021791912 | -0.106408135 |
| FinRL-A2C | -0.006822837 | -5.6857E-05 | 0.010206273 | -0.005570787 | 0.010306727 | -0.005516491 |
| FinRL-PPO | -0.046190532 | -0.000384921 | 0.013236838 | -0.029079534 | 0.013956025 | -0.027580998 |
| FinRL-SAC | -0.214470426 | -0.001787254 | 0.021616857 | -0.082678695 | 0.02253918 | -0.079295413 |
| FinRL-TD3 | -0.172235872 | -0.001435299 | 0.018900327 | -0.075940427 | 0.019436853 | -0.073844205 |
| EIIE | -0.376168975 | -0.003134741 | 0.03459112 | -0.090622722 | 0.035390026 | -0.088576976 |
| Dlinear | -0.256897969 | -0.002140816 | 0.022679636 | -0.094393771 | 0.023539744 | -0.09094476 |
| Autoformer | -0.32720313 | -0.002726693 | 0.019926248 | -0.136839245 | 0.020621049 | -0.132228616 |

Table 5: Empirical results in experiment 3. The metrics in this table are consistent with those in Table 3. Transaction costs and lending costs are included in buying and selling actions. The best results for the metric regarding the return and return per unit of risk are highlighted in bold.

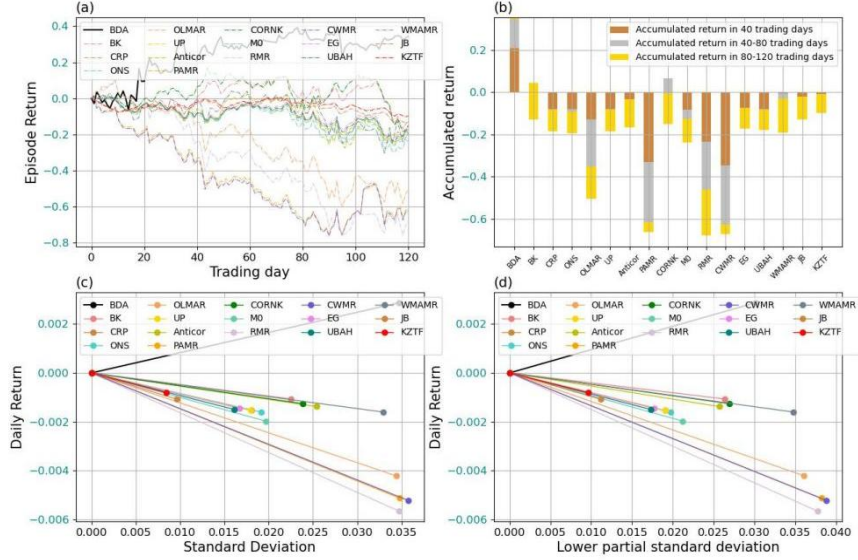

Fig 12. The visualization of the empirical results in Experiment 3. The figure reflects the empirical results of our BDA and the traditional comparative strategies.

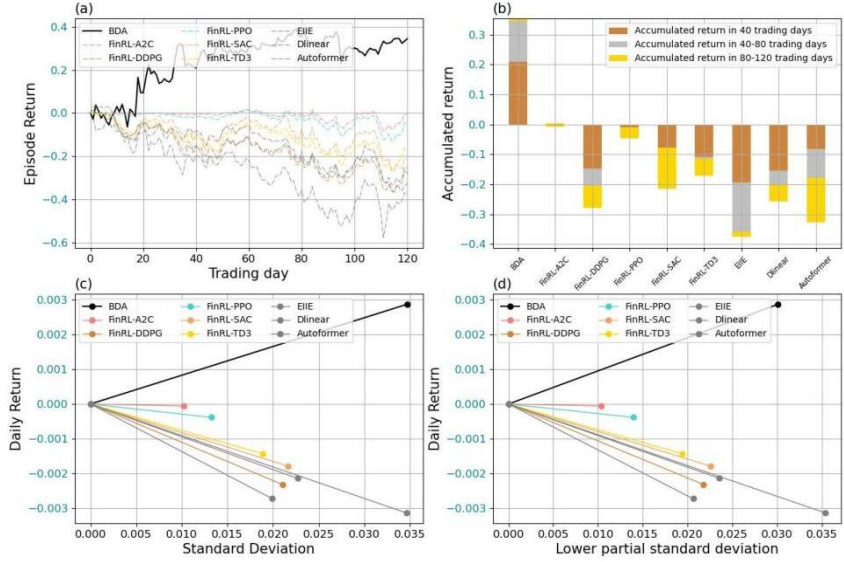

Fig 13. The visualization of the empirical results in Experiment 3. The figure reflects the empirical results of our BDA and the comparative strategies based on DL algorithms and DRL algorithms.

| Strategies | Performance Metrics | | | | | |
|---|---|---|---|---|---|---|
|  | AR | DR | Std | SR | LStd | STR |
| BDA | **0.343221341** | **0.002860178** | 0.035129119 | **0.081419003** | 0.035182984 | **0.08129435** |
| BK | 0.078257807 | 0.000652148 | 0.020880645 | 0.031232196 | 0.021307067 | 0.03060714 |
| CRP | 0.099331782 | 0.000827765 | 0.018458562 | 0.044844492 | 0.01781593 | 0.046462061 |
| ONS | 0.096635123 | 0.000805293 | 0.018605726 | 0.043281981 | 0.017758604 | 0.045346622 |
| OLMAR | 0.165244184 | 0.001377035 | 0.035316643 | 0.038991103 | 0.031751376 | 0.043369297 |
| UP | 0.099163803 | 0.000826365 | 0.01844314 | 0.044806092 | 0.017805472 | 0.046410734 |
| Anticor | 0.113945769 | 0.000949548 | 0.024012616 | 0.039543716 | 0.021731016 | 0.043695521 |
| PAMR | -0.592659823 | -0.004938832 | 0.034760484 | -0.142081791 | 0.037236573 | -0.132633899 |
| CORNK | 0.166133607 | 0.001384447 | 0.022619983 | 0.061204587 | 0.022495611 | 0.061542971 |
| M0 | 0.100310572 | 0.000835921 | 0.019794483 | 0.04223002 | 0.018478636 | 0.045237184 |
| RMR | 0.013517573 | 0.000112646 | 0.037914109 | 0.002971096 | 0.037286624 | 0.003021095 |
| CWMR | -0.612287416 | -0.005102395 | 0.035291904 | -0.144576929 | 0.037595366 | -0.135718724 |
| EG | 0.09750823 | 0.000812569 | 0.018410661 | 0.044135763 | 0.017968757 | 0.04522119 |
| UBAH | 0.08677255 | 0.000723105 | 0.018641823 | 0.038789372 | 0.018398643 | 0.039302061 |
| WMAMR | 0.091801637 | 0.000765014 | 0.032681185 | 0.023408382 | 0.034146569 | 0.022403822 |
| JB | -0.018918467 | -0.000157654 | 0.007633176 | -0.020653774 | 0.008786004 | -0.017943754 |
| KZTF | -0.008451192 | -7.04266E-05 | 0.006701401 | -0.010509235 | 0.007648304 | -0.009208133 |
| FinRL-DDPG | 0.083451918 | 0.000695433 | 0.020831188 | 0.033384205 | 0.01866029 | 0.037268052 |
| FinRL-A2C | 0.031667399 | 0.000263895 | 0.009517991 | 0.027725913 | 0.008921119 | 0.029580928 |
| FinRL-PPO | 0.038681627 | 0.000322347 | 0.012470647 | 0.025848449 | 0.011534983 | 0.027945156 |
| FinRL-SAC | 0.096154461 | 0.000801287 | 0.019192868 | 0.041749215 | 0.017726486 | 0.045202822 |
| FinRL-TD3 | 0.066808291 | 0.000556736 | 0.01443212 | 0.038576158 | 0.013495973 | 0.041251991 |
| EIIE | 0.240183926 | 0.002001533 | 0.025514117 | 0.078448048 | 0.023848562 | 0.083926768 |
| Dlinear | 0.119329953 | 0.000994416 | 0.02243819 | 0.044318026 | 0.022171571 | 0.044850962 |
| Autoformer | 0.123590284 | 0.001029919 | 0.016858357 | 0.061092491 | 0.017432466 | 0.059080513 |

Table 6: Empirical results in experiment 4. The metrics in this table are consistent with those in Table 3. Transaction costs and lending costs are included in buying and selling actions. The best results for the metric regarding the return and return per unit of risk are highlighted in bold.

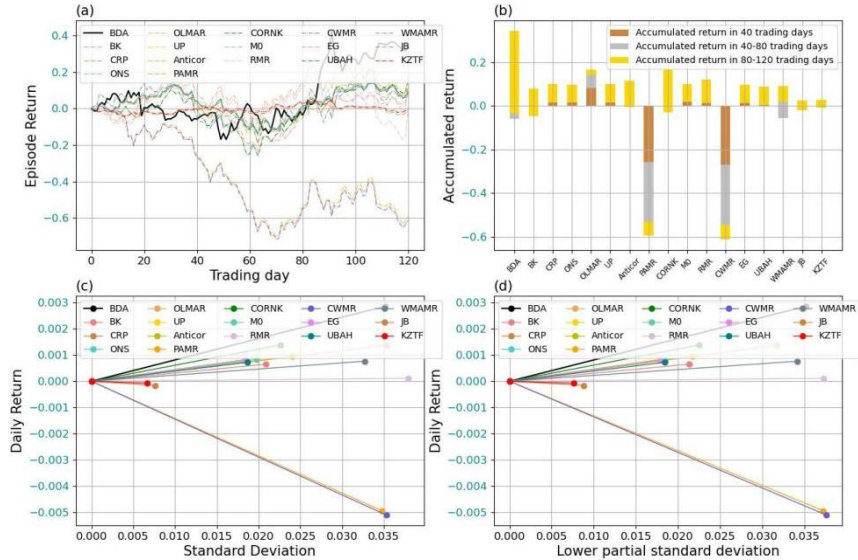

Fig 14. The visualization of the empirical results in Experiment 4. The figure reflects the empirical results of our BDA and the traditional comparative strategies.

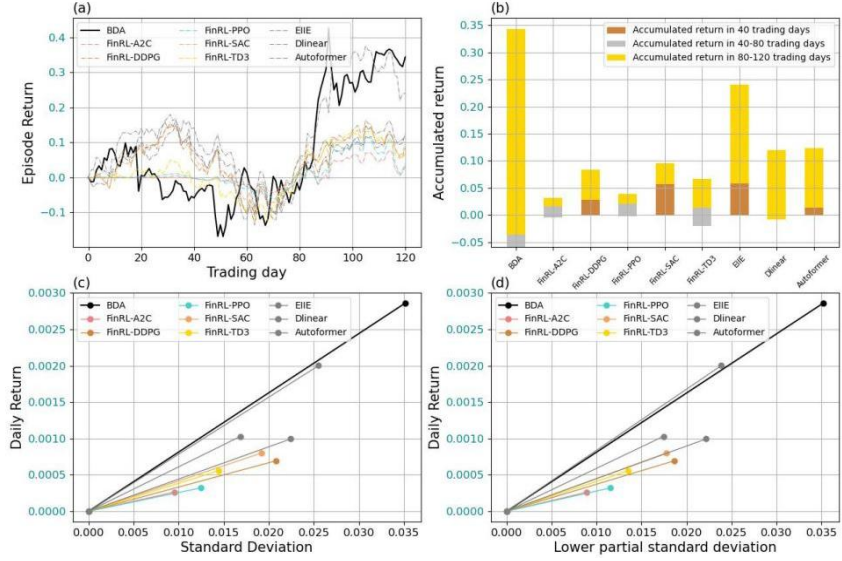

Fig 15. The visualization of the empirical results in Experiment 4. The figure reflects the empirical results of our BDA and the comparative strategies based on DL algorithms and DRL algorithms.

### 6.6 Ablation study

**The performance gain of the modified Transformer**

In our research, our BDA utilizes a Transformer network from which the position encoding module is removed. Utilizing this method enables our BDA to concentrate on learning the correlation among portfolio asset returns when determining its subjective views of the expected return. To verify the value of removing the position encoding module in ensuring the generalization ability of our DRL agent's policy, we compare our BDA with two variants (i.e. BDA-V1, BDA-V2). In the framework of the first variant, our modified transformer network is replaced by the original Transformer architecture, which contains the position encoding module. Furthermore, our transformer network also compares to the advanced conventional neural network structure. In the framework of the second variant, the Transformer network applied for determining the subjective views of the expected return is replaced by the CNN-LSTM-ResNet [82] structure. Table 7 gives the out-sample performance of our BDA and its two variants. The empirical results show that our BDA outperform the variants by at least 247% in terms of AR. In terms of SR and STR, our BDA outperforms the variants by at least 24% in all experiments. It demonstrates that the decision-making model concentrating on the correlation of asset returns can realize the generalization ability of our BDA's policy.

**The setting of the target value in the training process**

To address the issue of overfitting and realize the generalization ability of our BDA's policy function, we establish the target value for the evaluation function during the training process. To verify the value of the training target value in avoiding overfitting, we compare our BDA with a variant of BDA, which is trained by maximizing the evaluation function. From Table 8, we can observe that removing the application of the target value leads to a remarkable drop in the out-sample performance in AR and return per unit of risk. Specifically, the drop in SR and STR is at least 171% in all experiments. In terms of AR, removing the application of the target value leads to our BDA suffering loss in experiments 1 and 2. The empirical results reflect that setting the target value can effectively avoid overfitting and realize the generalization ability of our BDA's policy function in the training process.

**Necessity of the BL model**

Compared to previous attempts at applying DRL to portfolio management in recent years, i.e., [15, 23, 25, 26, 30, 32, 84, 85], the main novelty of this research is that we use the BL model to determine the portfolio weights instead of the softmax function used in these algorithms. To assess the effectiveness of the BL model in the portfolio decision process, we propose a variant of BDA in which the softmax activation function is applied to decide the target portfolio weights instead of the BL model. The policy function in the variant is expressed as

$$a_t = w_t = Softmax(N_1(s_t, \phi_1)).$$

As shown in Table 9, BDA exhibits better performance in accumulated return and return per unit of risk. Specifically, our BDA outperforms the variant of BDA (BDA-V4) by at least 246% in terms of AR. In terms of SR and STR, our BDA outperforms the variant of BDA (BDA-V4) by at least 7%. The empirical results indicate that implementing the long/short strategy guided by the BL model can further improve the agent's profitability per unit of risk assumed when the target markets permit short selling.

**Necessity of decreasing the trading frequency**

To prevent the DRL agent from overfitting the noise from the environment, we extend the trading period to decrease the trading frequency. To prove the necessity of decreasing the trading period, we propose a variant of BDA, which is trained to learn the policy for portfolio optimization in the one-day trading period. Table 10 shows that BDA significantly outperforms the variant by at least 12% in SR and STR. In terms of AR, although the variant is slightly higher than BDA in experiment 2. In all the experiments, the AR of the variant is not stable. In experiment 1, the variant suffers loss. The empirical results indicate that extending the trading period to five days is necessary. In this way, our BDA can perform better in the out-sample accumulated return and return per unit of risk.

| Experiment | Strategies | Performance Matrix | | | | | |
|---|---|---|---|---|---|---|---|
| | | AR= | DR | Std | SR | LStd | STR |
| 1 | BDA | **0.397118421** | **0.00330932** | 0.023735829 | **0.139422985** | 0.02380215 | **0.139034506** |
| | BDA-V1 | -0.050446459 | -0.000420387 | 0.019388563 | -0.021682223 | 0.020915264 | -0.020099538 |
| | BDA-V2 | -0.231048966 | -0.001925408 | 0.014836926 | -0.129771362 | 0.014815177 | -0.129961865 |
| 2 | BDA | **0.274821202** | **0.002290177** | 0.026187598 | **0.087452722** | 0.024067521 | **0.09515632** |
| | BDA-V1 | 0.079065775 | 0.000658881 | 0.021508315 | 0.030633801 | 0.022475974 | 0.029314923 |
| | BDA-V2 | 0.04131821 | 0.000344318 | 0.007874372 | 0.043726462 | 0.007138046 | 0.04823707 |
| 3 | BDA | **0.344678392** | **0.00287232** | 0.034691237 | **0.082796701** | 0.030119603 | **0.095363804** |
| | BDA-V1 | -0.383470955 | -0.003195591 | 0.039313414 | -0.08128501 | 0.037767473 | -0.084612261 |
| | BDA-V2 | -0.194014737 | -0.001616789 | 0.015018357 | -0.107654218 | 0.015715172 | -0.1028808 |
| 4 | BDA | **0.343221341** | **0.002860178** | 0.035129119 | **0.081419003** | 0.035182984 | **0.08129435** |
| | BDA-V1 | -0.037574331 | -0.000313119 | 0.021081801 | -0.014852594 | 0.022033441 | -0.0142111 |
| | BDA-V2 | 0.073551653 | 0.00061293 | 0.011277665 | 0.054349057 | 0.00937014 | 0.065413159 |

Table 7: The ablation study of the proposed modified Transformer from which the position encoding module is removed. In the table, the variant applying the Transformer network with the position encoding module is termed BDA-V1. The variant applying CNN-ResNet-Lstm network structure is termed BDA-V2. The ablation experiments are conducted in the framework of four experiments presented in Table 1. The metrics in this table are consistent with those in Table 3. The better results for the metric regarding the return and return per unit of risk are highlighted in bold.

| Experiment | Strategies | Performance Matrix | | | | | |
|---|---|---|---|---|---|---|---|
| | | AR | DR | Std | SR | LStd | STR |
| 1 | BDA | **0.397118421** | **0.00330932** | 0.023735829 | **0.139422985** | 0.02380215 | **0.139034506** |
| | BDA-V3 | -0.757258438 | -0.006310487 | 0.073384174 | -0.085992478 | 0.07882949 | -0.080052363 |
| 2 | BDA | **0.274821202** | **0.002290177** | 0.026187598 | **0.087452722** | 0.024067521 | **0.09515632** |
| | BDA-V3 | -2.121894994 | -0.017682458 | 0.120113487 | -0.147214594 | 0.139773412 | -0.126508025 |
| 3 | BDA | **0.344678392** | **0.00287232** | 0.034691237 | **0.082796701** | 0.030119603 | **0.095363804** |
| | BDA-V3 | 0.25271908 | 0.002105992 | 0.095705494 | 0.022004926 | 0.099026566 | 0.021266943 |
| 4 | BDA | **0.343221341** | **0.002860178** | 0.035129119 | **0.081419003** | 0.035182984 | **0.08129435** |
| | BDA-V3 | 0.271695666 | 0.002264131 | 0.081540993 | 0.027766777 | 0.075540525 | 0.029972396 |

Table 8: The ablation study concerning the setting of the target for the evaluation function in the training process. The variant trained by maximizing the evaluation function is termed BDA-V3. The ablation experiments are conducted in the framework of four experiments presented in Table 1. The metrics in this table are consistent with those in Table 3. The better results for the metric regarding the return and return per unit of risk are highlighted in bold.

| Experiment | Strategies | Performance Matrix | | | | | |
|---|---|---|---|---|---|---|---|
| | | AR | DR | Std | SR | LStd | STR |
| 1 | BDA | **0.397118421** | **0.00330932** | 0.023735829 | **0.139422985** | 0.02380215 | **0.139034506** |
| | BDA-V4 | 0.114614775 | 0.000955123 | 0.007678048 | 0.124396604 | 0.007373812 | 0.12952909 |
| 2 | BDA | **0.274821202** | **0.002290177** | 0.026187598 | **0.087452722** | 0.024067521 | **0.09515632** |
| | BDA-V4 | 0.052009308 | 0.000433411 | 0.008832437 | 0.049070366 | 0.010389688 | 0.041715488 |
| 3 | BDA | **0.344678392** | **0.00287232** | 0.034691237 | **0.082796701** | 0.030119603 | **0.095363804** |
| | BDA-V4 | -0.076659093 | -0.000638826 | 0.019788445 | -0.032282768 | 0.020537432 | -0.031105436 |
| 4 | BDA | **0.343221341** | **0.002860178** | 0.035129119 | **0.081419003** | 0.035182984 | **0.08129435** |
| | BDA-V4 | 0.007721546 | 6.43462E-05 | 0.007370779 | 0.008729907 | 0.007963335 | 0.00808031 |

Table 9: The ablation study of the BL model. In the table, the variant in which the softmax activation function is applied to decide the target portfolio weights instead of the BL model is termed BDA-V4. The ablation experiments are conducted in the framework of four experiments presented in Table 1. The metrics in this table are consistent with those in Table 3. The better results for the metric regarding the return and return per unit of risk are highlighted in bold.

| Experiment | Strategies | Performance Matrix | | | | | |
|---|---|---|---|---|---|---|---|
| | | AR | DR | Std | SR | LStd | STR |
| 1 | BDA | **0.397118421** | **0.00330932** | 0.023735829 | **0.139422985** | 0.02380215 | **0.139034506** |
|   | BDA-V5 | -1.406088638 | -0.011717405 | 0.186282163 | -0.062901381 | 0.199602602 | -0.05870367 |
| 2 | BDA    | 0.274821202  | 0.002290177  | 0.026187598 | **0.087452722** | 0.024067521 | **0.09515632** |
|   | BDA-V5 | **0.357159214** | **0.002976327** | 0.04027183 | 0.073905923 | 0.035296236 | 0.084324197 |
| 3 | BDA    | **0.344678392** | **0.00287232** | 0.034691237 | **0.082796701** | 0.030119603 | **0.095363804** |
|   | BDA-V5 | 0.170726813  | 0.001422723  | 0.069070457 | 0.020598147 | 0.072821627 | 0.019537101 |
| 4 | BDA    | **0.343221341** | **0.002860178** | 0.035129119 | **0.081419003** | 0.035182984 | **0.08129435** |
|   | BDA-V5 | 0.303559825  | 0.002488195  | 0.076179212 | 0.032662392 | 0.071898211 | 0.034607193 |

Table 10: The ablation study of extending the trading period. In the table, the variant of BDA, which learns the policy for portfolio optimization in a one-day trading period, is termed as BDA-V5. The ablation experiments are conducted in the framework of four experiments presented in Table 1. The metrics in this table are consistent with those in Table 3. The better results for the metric regarding the return and return per unit of risk are highlighted in bold.

## 7. Conclusions and future work

This paper proposes a DRL framework for the target market where short selling is permitted. In our proposed framework, the DRL agent is trained to learn the policy to apply the mathematical BL model to determine target portfolio weights in consecutive trading periods. The out-sample empirical results show that, by training the DRL agent to learn the policy to apply the BL model for portfolio optimization, the DRL agent has outstanding performance in accumulated return and the return per unit of risk. It demonstrates that learning the policy to apply the BL model is a feasible method for the DRL agent to achieve outstanding performance in return per unit of risk while maintaining outstanding profitability in the short permitting market. To ensure the effective realization of our DRL agent's training objective, we choose to train our DRL agent by formulating the objective function based on the reward function in the environment and propagating the gradient of the objective function to the policy function of our DRL agent. The in-sample empirical results prove that, in the portfolio optimization problem with continuous and high dimensional action space, our training method is more effective than the traditional actor-critic algorithm for deterministic policy in maximizing the DRL agent's accumulative rewards. To realize the generalization ability of our DRL agent, we remove the position encoding module in the Transformer network to let our BDA concentrate on learning to extract non-linear correlation from multiple concurrent time series of portfolio assets when deriving expectations of returns. Meanwhile, for the function derived from the environment's reward function, we establish the training target values instead of maximizing such function in the training process to avoid overfitting. From the out-sample ablation study, we can conclude that these methods play a crucial role in ensuring the generalization ability of our DRL agent's policy.

Despite these promising results, there are two major limitations in our model. First, the BL model [44] assumes that the portfolio asset returns follow a Gaussian distribution. The distribution of the asset returns may exhibit heavier tails and occasionally high peaks in reality [83], and the Gaussian distribution cannot describe these features. Hence, the DRL agent cannot effectively capture all the characters in the historical return distribution. Second, in terms of accumulative return and Sharpe ratio, the performance of the DRL agent in the back-test environments is inferior to its performance in the training environments. This implies that we need to further improve the DRL agent's ability to generalize its learned behaviour from the training environments to the back-test environments. For future work, we shall investigate how the above drawbacks may be overcome to further improve the performance of the DRL agent. First, we plan to train the DRL agent to learn the policy to apply the Bayesian model based on the Elliptical distributions. This way, the DRL agent can effectively describe the heavier tails and occasionally high peak features in the return distribution. To further improve the generalization ability of the DRL agent's portfolio optimization policy, we will construct a multi-agent framework and apply hierarchical reinforcement learning in future work.

**Data availability**

This research uses only publicly available American stock data from Yahoo Finance platform. All datasets generated during and/or analyzed during the current study, models, or codes that support the findings of this study are available from the corresponding author upon reasonable request.

**Declarations**

**Conflict of interest** The authors declare that they have no known competing financial interests or personal relationships that could have appeared to influence the work reported in this paper.

# Appendix 1

Hyper-parameters in the paper

| Where | Hyper-parameters | Value |
|---|---|---|
| Hyper-parameters in the state tensor | Number of trading periods $m$ included in the historical return tensor | 50 |
| Hyper-parameters in the portfolio | Number of stocks $n$ included in the historical return tensor | 29 |
| Hyper-parameters in the portfolio | The annual lending rate $r_l$ | 0.03 |
| Hyper-parameters in the BL model | Scalar $\tau$ describing the confidence level of the prior expectation | 1 |
| Hyper-parameters in the reward function | The parameters $\lambda_1$ to describe the risk aversion in the elevation function | 0.2 |
| Hyper-parameters in the reward function | The parameters $\lambda_2$ to limit the transaction scale in the elevation function | 0.002 |
| Hyper-parameters in the reward function | the risk aversion $\lambda_3$ in the calculation of the optimal portfolio weights | 1 |
| Hyper-parameters in the DRL training | Target step $M$ | 1080 |
| Hyper-parameters in the DRL training | Minibatch size $N$ | 128 |
| Hyper-parameters in the DRL training | Learning rate $\alpha$ | 1e-5 |
| Hyper-parameters in the DRL training | Reply buffer size | 2^14 |
| Hyper-parameters in the neural network | Depth of the Transformer encoder block $L$ | 6 |
| Hyper-parameters in the neural network | The size of the queries $q$, keys $k$, and values $v$: $L_{model}$ | 29 |
| Hyper-parameters in the neural network | Value $d$ is the scaling factor | 1 |
| Hyper-parameters in the neural network | Value $l$ in the fully-connected network $W_1$ and $W_2$ | 3712 |

# Appendix 2

The list of the Dow Jones Industrial Average (DJIA) components for portfolio construction and their respective tickers, names, and categories

| No. | Ticker | Name | Category |
|---|---|---|---|
| 1 | MMM | 3M Company | Industrial |
| 2 | AXP | American Express Company | Financial |
| 3 | AMGN | Amgen Inc. | Healthcare |
| 4 | AAPL | Apple Inc. | Technology |
| 5 | BA | Boeing Company | Industrial |
| 6 | CAT | Caterpillar Inc. | Industrial |
| 7 | CVX | Chevron Corporation | Energy |
| 8 | CSCO | Cisco Systems, Inc. | Technology |
| 9 | KO | Coca-Cola Company | Consumer goods |
| 10 | GS | Goldman Sachs Group, Inc. | Financial |
| 11 | HD | The Home Depot, Inc. | Consumer services |
| 12 | HON | Honeywell International Inc. | Industrial |
| 13 | IBM | IBM Corporation | Technology |
| 14 | INTC | Intel Corporation | Technology |
| 15 | JNJ | Johnson & Johnson | Healthcare |
| 16 | JPM | JPMorgan Chase & Co. | Financial |
| 17 | MCD | McDonald's Corporation | Consumer services |
| 18 | MRK | Merck & Co., Inc. | Healthcare |
| 19 | MSFT | Microsoft Corporation | Technology |
| 20 | NKE | Nike, Inc. | Consumer goods |
| 21 | PG | Procter & Gamble Company | Consumer goods |
| 22 | CRM | Salesforce.com, Inc. | Technology |
| 23 | TRV | The Travelers Companies, Inc. | Financial |
| 24 | UNH | UnitedHealth Group Incorporated | Healthcare |
| 25 | VZ | Verizon Communications Inc. | Communication |
| 26 | V | Visa Inc. | Financial |
| 27 | WBA | Walgreens Boots Alliance, Inc. | Consumer goods |
| 28 | WMT | Walmart Inc. | Consumer services |
| 29 | DIS | The Walt Disney Company | Consumer services |